\documentclass[aps,prc,twocolumn,showpacs,superscriptaddress,footinbib]{revtex4-1}

\usepackage{graphicx}
\usepackage{dcolumn}
\usepackage{bm}
\usepackage{enumerate}
\usepackage{amsmath}

\usepackage[normalem]{ulem}  

\begin{document}


\title{Level-anticrossing in B(E2) anomaly (I)}

\author{Tao Wang}
\email{suiyueqiaoqiao@163.com}
\affiliation{College of Physics, Tonghua Normal University, Tonghua 134000, People's Republic of China}

\author{Yu-xin Cheng}
\affiliation{School of Physics, Liaoning University, Shenyang 110036, People's Republic of China}

\author{Dong-kang Li}
\email{ldk667788@163.com}
\affiliation{College of Physics, Tonghua Normal University, Tonghua 134000, People's Republic of China}

\author{Xiao-shen Kang}
\email{kangxiaoshen@lnu.edu.cn}
\affiliation{School of Physics, Liaoning University, Shenyang 110036, People's Republic of China}

\author{Suo-chang Jin}
\affiliation{Department of Physics, College of Science, Yanbian University, Yanji, Jilin 133002, China}

\author{Tie Wang}
\email{twang@ybu.edu.cn}
\affiliation{Department of Physics, College of Science, Yanbian University, Yanji, Jilin 133002, China}

\author{Zhi-qi Zhang}
\affiliation{Department of Physics, Liaoning Normal University, Dalian 116029, People's Republic of China}

\author{Chen-guang Zhang}
\affiliation{College of Physics, Tonghua Normal University, Tonghua 134000, People's Republic of China}

\author{Zhi-xin Zhang}
\affiliation{College of Physics, Tonghua Normal University, Tonghua 134000, People's Republic of China}

\date{\today}

\begin{abstract}
Recently, a new mechanism for explaining the B(E2) anomaly was given by F. Pan \emph{et al.} (PRC, 110, 054324, 2024), which is realized in the parameter region from the SU(3) symmetry limit to the O(6) symmetry limit, and seems to be not related to the SU(3) symmetry. However, through SU(3) analysis, a new technique proposed recently, we found that it is not so. The new mechanism is related to level-anticrossing phenomenon, which is related to level-crossing phenomenon in the SU(3) symmetry limit. By incorporating previous ideas, we have a more general explanatory framework for the B(E2) anomaly, which is important for understanding some higher-order interactions in the interacting boson model. Through analysis, it is shown that level-anticrossing in this mechanism mainly results from the third-order interaction $[\hat{L}\times \hat{Q}_{\chi} \times \hat{L}]^{(0)}$. Finally, the B(E2) anomaly in $^{170}$Os is also discussed within this general framework.
\end{abstract}

\maketitle

\section{Introduction}

50 years ago, the interacting boson model (IBM) was proposed by Arima and Iachello \cite{Iachello75,Iachello87}, which is a widely influential algebraic model for describing the collective behaviors of various nuclei. In the simplest model, the $s$ bosons ($L=0$) and $d$ bosons ($L=2$) are considered, which has the U(6) symmetry. Four dynamical symmetric limit exist: (1) the U(5) symmetry limit can describe the spherical vibration, (2) the SU(3) symmetry limit can present the prolate rotation, (3) the O(6) symmetry limit can describe the $\gamma$-soft rotation, and (4) the $\overline{\textrm{SU(3)}}$ symmetry limit can present the oblate rotation \cite{Jolie01}. In this model, shape phase transitions from the spherical shape to various quadrupole deformations or among the deformed shapes can be studied \cite{Warner02,Casten06,Casten07,Bonatsos09,Casten09,Jolie09,Casten10,Jolos21,Fortunato21,Cejnar21,Jolie00,Cejnar03,Iachello04,Wang08}. A particularly interesting scenario is the shape phase transition from the prolate shape to the oblate shape \cite{Bonatsos24}, where the O(6) symmetry limit is also the first-order phase transitional point \cite{Jolie01}. In this description, the energy spectra of the prolate and oblate shapes are the same for the same boson number $N$ \cite{Wang08}. However this mirror symmetry is not found in realistic nuclei. In Ref. \cite{Jolie03}, the energy ratio $E_{4/2}=E_{4_{1}^{+}}/E_{2_{1}^{+}}$ of the $4_{1}^{+}$ and $2_{1}^{+}$ states of the realistic nuclei in the Hf-Hg region is 3.33 for the prolate shape, while it is 2.55 for the oblate shape. Clearly they are asymmetric.

Although previous IBM did give many good descriptions of the low-energy collective excitations of nuclei, some recent new experiments have been found to conflict with the results of previous IBM and other nuclear structure theories. Among these anomalous phenomena, the most prominent one is the B(E2) anomaly \cite{Grahn16,Saygi17,Cederwall18,Goasduff19,Stolze21}. In general cases, the E2 transition ratio $B_{4/2}=B(E2;4_{1}^{+}\rightarrow 2_{1}^{+})/B(E2;2_{1}^{+}\rightarrow 0_{1}^{+})$ among the $0_{1}^{+}$, $2_{1}^{+}$ and $4_{1}^{+}$ states is usually larger than 1.0 for collective excitations while the ratio $E_{4/2}$ is larger than 2.0. All the time, this seems to be the real case. B(E2) anomaly found in $^{168,170}$Os \cite{Grahn16,Goasduff19}, $^{166}$W \cite{Saygi17} and $^{172}$Pt \cite{Cederwall18} changes this old idea. In these nuclei, the $B_{4/2}$ value can be smaller than 1.0, even reduces to 0.33, while the $E_{4/2}$ value is still larger than 2.0. The discovery of the B(E2) anomaly has surprised the nuclear structure researchers, but so far, this phenomenon can not be explained by other nuclear structure theories. Since this anomaly only involves the three $0_{1}^{+}$, $2_{1}^{+}$ and $4_{1}^{+}$ states, explaining it is a very important thing for any nuclear structure model, and it can be regarded as a criterion to judge whether a model is reasonable. Recently, B(E2) anomaly was also found in the odd-even nuclei $^{167,169}$Os \cite{Cederwall21,Zanon25}.

Recently an extension of the interacting boson model with SU(3) higher-order interactions (SU3-IBM) was proposed by one of the authors (T. Wang) \cite{Wang20,Wang22}. This new model combines the idea of the IBM \cite{Iachello87} and the SU(3) correspondence of the rigid triaxial rotor \cite{Isacker85,Isacker00,Draayer87,Draayer881,Kota20}. The SU(3) symmetry dominates all the quadrupole deformations (prolate, oblate and various rigid triaxial). Clearly for the description of the oblate shape, the SU3-IBM is very different from previous IBM. In the SU3-IBM, only the U(5) symmetry limit and the SU(3) symmetry limit are included. In the SU(3) symmetry limit, the SU(3) second-order Casimir operator $-\hat{C}_{2}[SU(3)]$ can describe the prolate shape, and the SU(3) third-order Casimir operator $\hat{C}_{3}[SU(3)]$ can present the oblate shape. Moreover, when the square of the second-order Casimir operator $\hat{C}_{2}^{2}[SU(3)]$ is added into previous two ones, the rigid triaxial deformations with any SU(3) irrep $(\lambda,\mu)$ can be described. Other dynamical higher-order interactions $[\hat{L}\times \hat{Q} \times \hat{L}]^{(0)}$ and $[(\hat{L}\times \hat{Q})^{(1)}\times (\hat{L} \times \hat{Q})^{(1)}]^{(0)}$ are also needed ($\hat{Q}$ is the SU(3) quadrupole operator). The B(E2) anomaly has been successfully explained by the SU3-IBM \cite{Wang20,Zhang22,Wangtao,Zhang24,Pan24,Zhang25,Zhang252,Cheng25}, but so far many possible mechanisms exist.

The SU3-IBM can also explain the Cd puzzle \cite{Wang22,Wang25,WangPd,Zhao251,Zhao252}. In this puzzle, experimental researchers found that the vibrational phonon excitations of the spherical nucleus, which has long been thought to exist, cannot be confirmed experimentally \cite{Garrett08,Garrett12,Batchelder12,Garrett19,Garrett20,Garrett18}. This seriously challenges the traditional view of the emergence of collectivity and shape evolutions in nuclear structure. One of the authors (T. Wang) argued that these experiments found a new collective excitation mode \cite{Wang22}. It was called  spherical-like nucleus and has been found in the SU3-IBM. The SU3-IBM was used to describe the whole $^{108-120}$Cd nuclei, and the anomalous evolutional trend of the electric quadrupole moment $Q_{2_{1}^{+}}$ of the first $2^{+}$ states in $^{108-116}$Cd was explained, in which the $Q_{2_{1}^{+}}$ value decreases while the boson number $N$ increases \cite{Wang25}. Recently $^{106}$Pd is found to be a typical spherical-like nucleus \cite{WangPd}, which directly shows that this new spherical-like collectivity really exists. Moreover, the new $\gamma$-softness is found to be really a shape phase. The shape phase transition from the new $\gamma$-soft phase to the prolate shape really exits and $^{108}$Cd may be the critical nucleus \cite{Zhao251,Zhao252}.  The Cd puzzle and the B(E2) anomaly occur for small boson number $N\leq 9$, so they may have a common origin. The succuss of the SU3-IBM supports this idea.

The SU3-IBM can also explain the prolate-oblate asymmetric shape phase transition in the Hf-Hg region \cite{Wang23}, can describe the properties of $^{^{196}}$Pt at a better level \cite{WangPt,ZhouPt}, can describe the E(5)-like spectra in $^{82}$Kr \cite{Zhou23}. Recently, two important results are found. In the SU3-IBM, the oblate shape is described by the SU(3) third-order Casimir operator, thus within the SU(3) symmetry limit, there exists a new path for the prolate-oblate shape phase transition \cite{Fortunato11}, which is different from the SU(3)-$\overline{\textrm{SU(3)}}$ description in previous IBM \cite{Jolie01}. Then this prolate-oblate asymmetric shape phase transition was further studied by Y. Zhang \emph{et al.}, which predicts a unique boson number odd-even phenomenon for the $0_{2}^{+}$ and $2_{2}^{+}$ states for the oblate nuclei \cite{Zhang12}. One of the authors (T. Wang) and his coworkers really found it in $^{196-204}$Hg recently \cite{WangHg}. This unique phenomenon can not be explained by previous theories \cite{Bernards131,Bernards132}. Another important result is that, Otsuka \emph{et al.} proposed that, the large-deformed nuclei previously thought as the prolate shape, such as $^{166}$Er, should be rigid triaxial with around $\gamma=8^{\circ}$ \cite{Otsuka19,Otsuka21,Otsuka}. Recently they found that $^{154}$Sm is also a rigid triaxial nucleus \cite{Otsuka}. This rigid triaxiality can be also explained by the SU3-IBM \cite{ZhouEr}. Together these findings support that the SU3-IBM provides a better Hamiltonian to describe the collectivity of nuclei and that this model is valid and correct.

This implies that using the SU3-IBM to explain the B(E2) anomaly is credible, and warrants further investigations, especially on the mechanism of this phenomenon. In Ref. \cite{zhang14}, when investigating the SU(3) correspondence of the rigid triaxial rotor, Zhang \emph{et al.} found the B(E2) anomaly theoretically for the first time, but they did not believe this anomaly really exists in realistic nuclei. Inspired by the experimental findings of the B(E2) anomaly in $^{168,170}$Os, $^{166}$W and $^{172}$Pt, one of the authors (T. Wang) presented the first theoretical description \cite{Wang20}. In this explanation, the SU(3) third-order interaction $[\hat{L}\times \hat{Q} \times \hat{L}]^{(0)}$ plays a key role. In the SU(3) symmetry limit, the $[\hat{L}\times \hat{Q} \times \hat{L}]^{(0)}$ interaction can reduce the energy of one $4^{+}$ state in the SU(3) irrep ($2N-8,4$) and increases the energy of the $4^{+}$ state in the SU(3) irrep ($2N,0$), thus the former one can be lower than the latter one, and the ratio $B_{4/2}$ is zero, which is a possible origin of the B(E2) anomaly. For this occurs within the SU(3) symmetry limit and states with the same angular momentum and different SU(3) irreps are not correlated, it is a level-crossing phenomenon.

Subsequently, Y. Zhang \emph{et al.} used the SU(3) correspondence of the rigid triaxial rotor to explain the B(E2) anomaly \cite{Zhang22}. Thus two very different theories exist, but they are difficult to distinguish due to insufficient experimental data. It becomes more important to further explore these anomalous mechanisms.

A central question of these studies is whether the emergence of the B(E2) anomaly is related to the SU(3) symmetry. In Ref. \cite{Wangtao}, up to fourth-order interactions, the B(E2) anomaly can not be explained by the O(6) symmetry. In the SU(3) description in \cite{Wang20}, when the $[\hat{L} \times \hat{Q} \times \hat{L}]^{(0)}$ interaction increases, the $4_{1}^{+}$ state can crossover with one other $4^{+}$ state, but in the O(6) description, the same phenomenon can not occur. This is very beneficial for determining the mechanism of the B(E2) anomaly.

Recently, a deeper understanding on the B(E2) anomaly was obtained by F. Pan \emph{et al.} \cite{Pan24}, which investigates the possibility between the SU(3) symmetry limit and the O(6) symmetry limit. They found a new mechanism for the emergence of the B(E2) anomaly. In their paper, this new mechanism seems to be not related to the SU(3) symmetry because in the SU(3) symmetry limit, the $B_{4/2}$ value is larger than 1.0. It directly seems that, in this parameter region, B(E2) anomaly can appear alone.

This new result attracts us to further investigate it. In a recent paper, the concept ``SU(3) analysis" was proposed \cite{Cheng25}, which is a useful technique to discuss the relationship between the B(E2) anomaly and the SU(3) symmetry. In SU(3) analysis, only the corresponding SU(3) symmetry limit is considered if it really exists, and the low-lying levels and the E2 transitional rates are studied when the parameter in front of the $[\hat{L}\times \hat{Q} \times \hat{L}]^{(0)}$ interaction varies. Three new results has been found: (1) the third-order interaction $[\hat{L}\times \hat{Q} \times \hat{L}]^{(0)}$ is vital for the B(E2) anomaly, (2) level-crossing phenomenon is important for the B(E2) anomaly, and (3) not only the E2 transition $B(E2;4_{1}^{+}\rightarrow 2_{1}^{+})$ but also $B(E2;2_{1}^{+}\rightarrow 0_{1}^{+})$ and $B(E2;6_{1}^{+}\rightarrow 4_{1}^{+})$ can be anomalous.

\begin{figure}[tbh]
\includegraphics[scale=0.33]{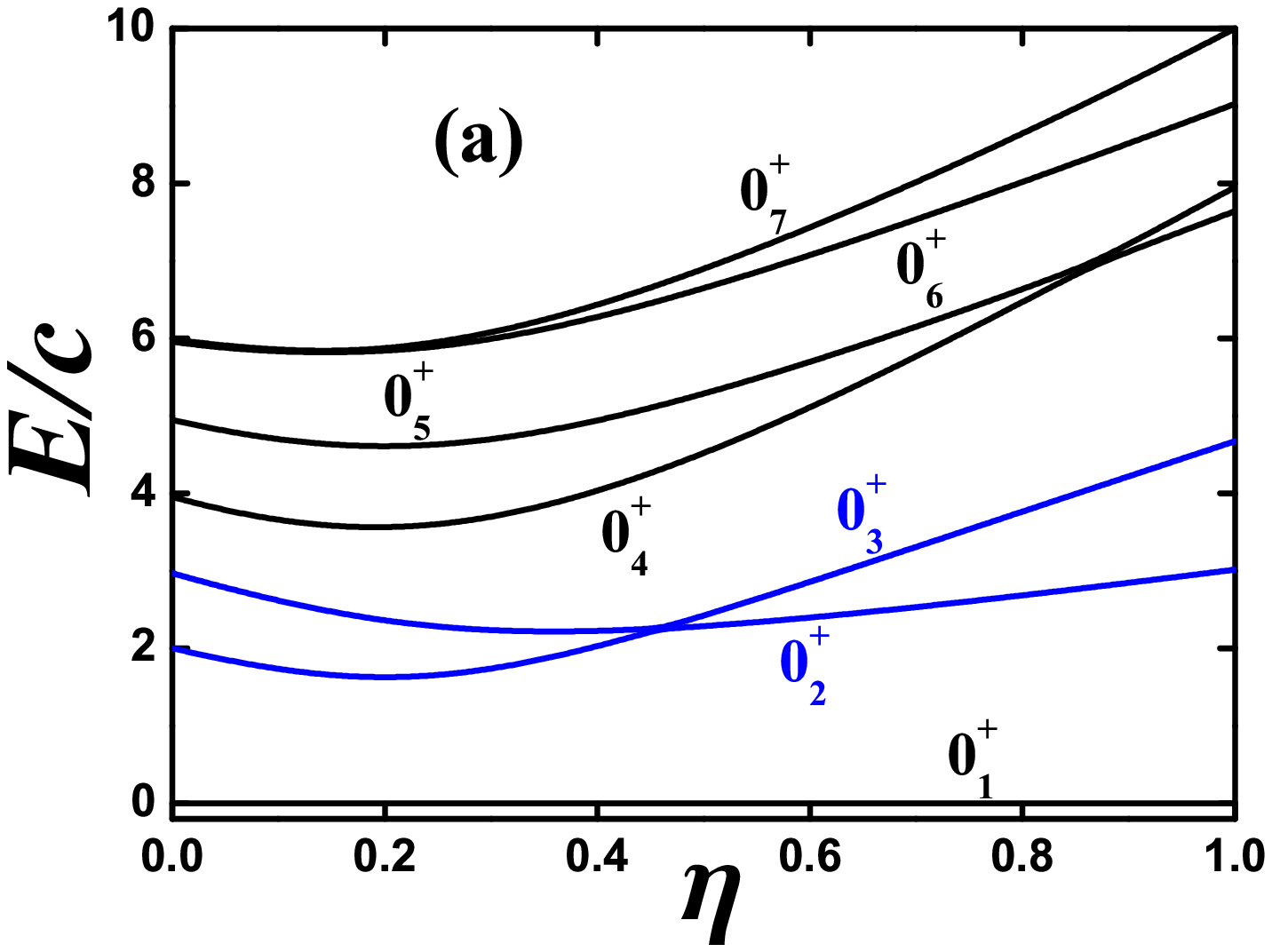}
\includegraphics[scale=0.33]{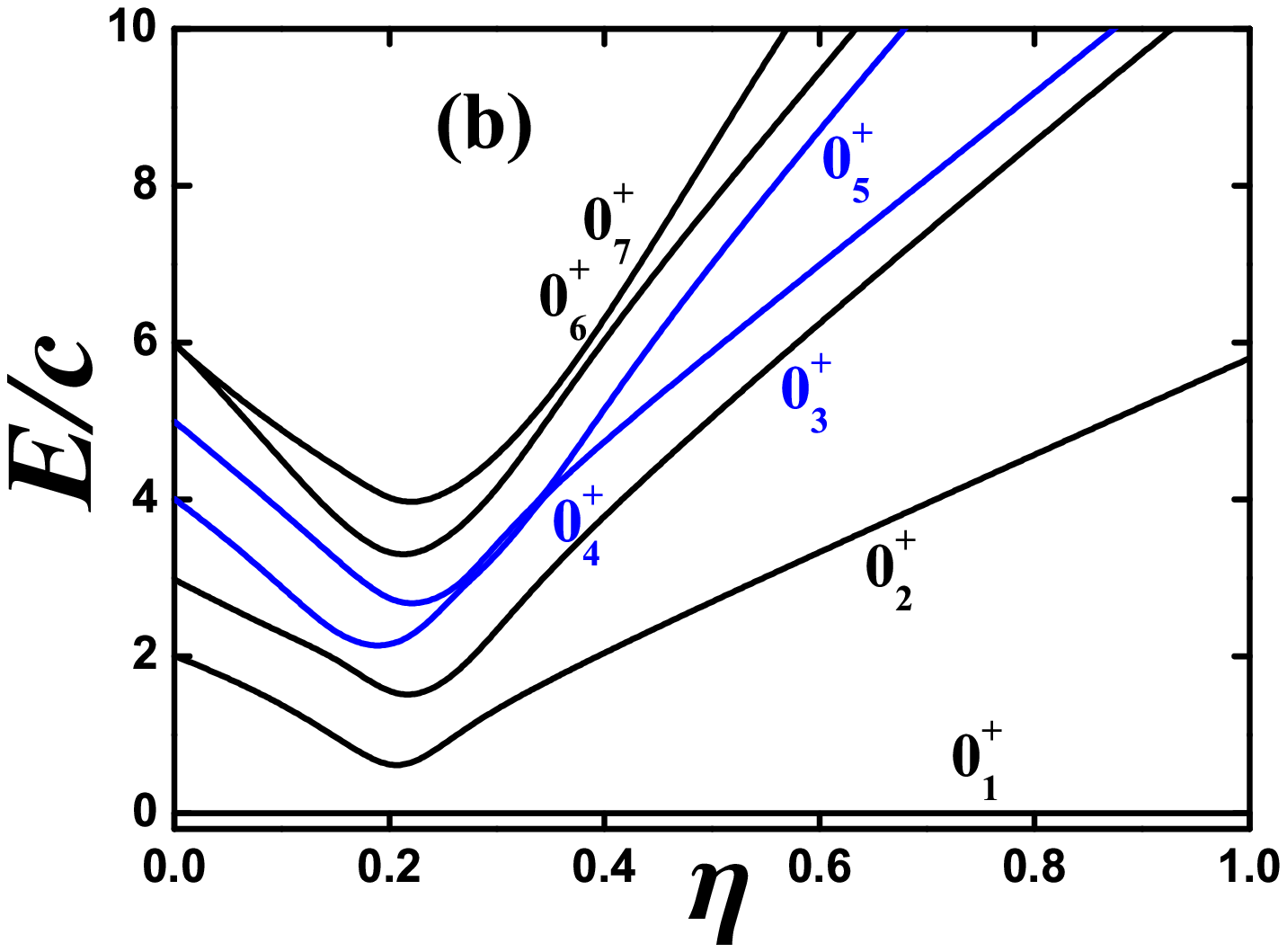}
\caption{(a) The evolutional behaviors of the $0^{+}$ states as a function of $\eta$ in $\hat{H}_{1}$ for $N=7$ from the U(5) symmetry limit to the O(6) symmetry limit; (b) The evolutional behaviors of the partial $0^{+}$ states as a function of $\eta$ in $\hat{H}_{2}$ for $N=15$ from the U(5) symmetry limit to the SU(3) symmetry limit.}
\end{figure}

We found the new mechanism in \cite{Pan24} is also related to the SU(3) symmetry. When the parameter of the $[\hat{L}\times \hat{Q} \times \hat{L}]^{(0)}$ interaction further decreases, level-crossing between the $4_{1}^{+}$ state and one other $4^{+}$ state can occur. However in that paper, the new mechanism in \cite{Pan24} is still unknown for that paper only discussed this problem within the SU(3) symmetry limit.

\begin{figure}[tbh]
\includegraphics[scale=0.33]{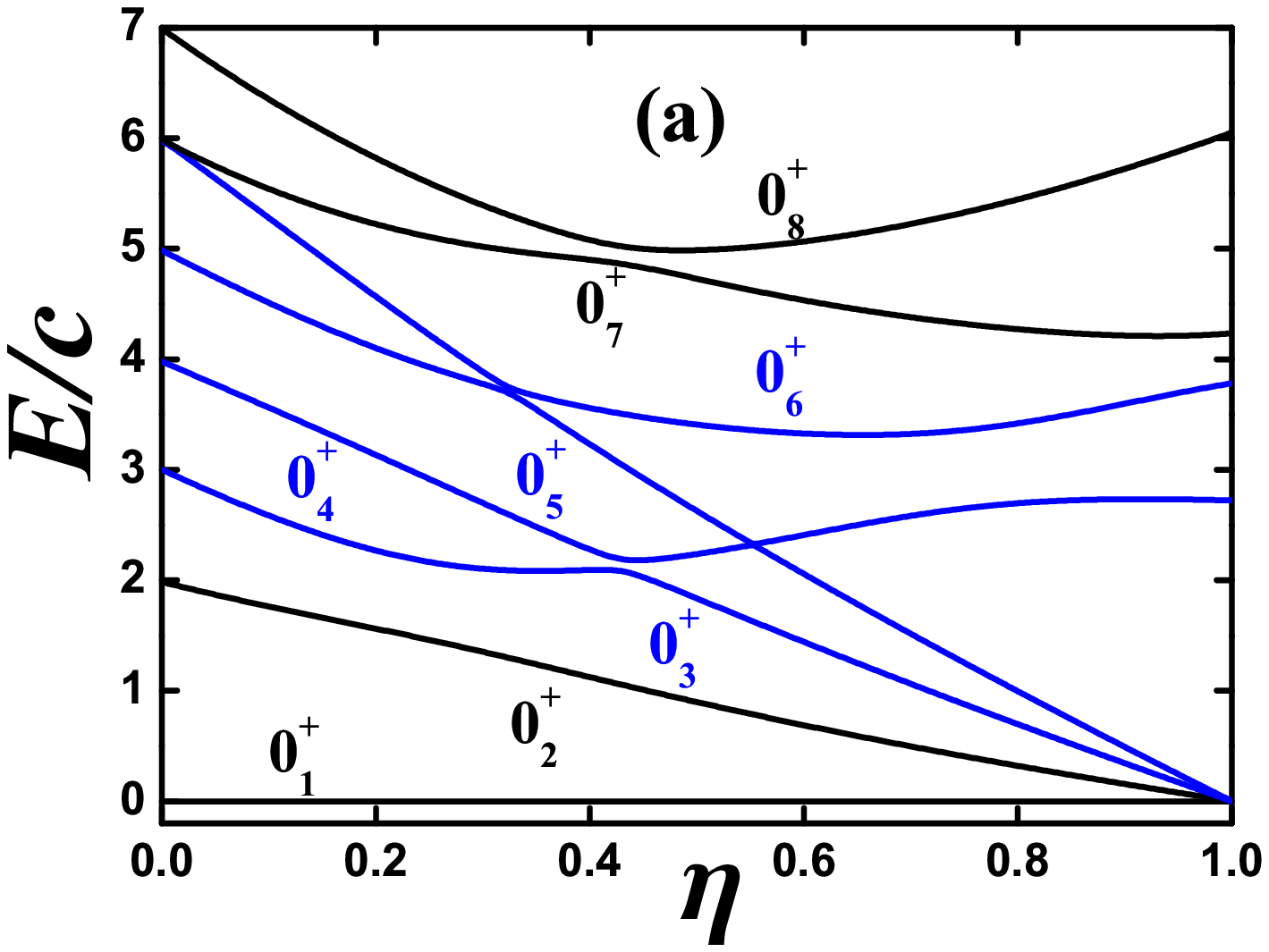}
\includegraphics[scale=0.33]{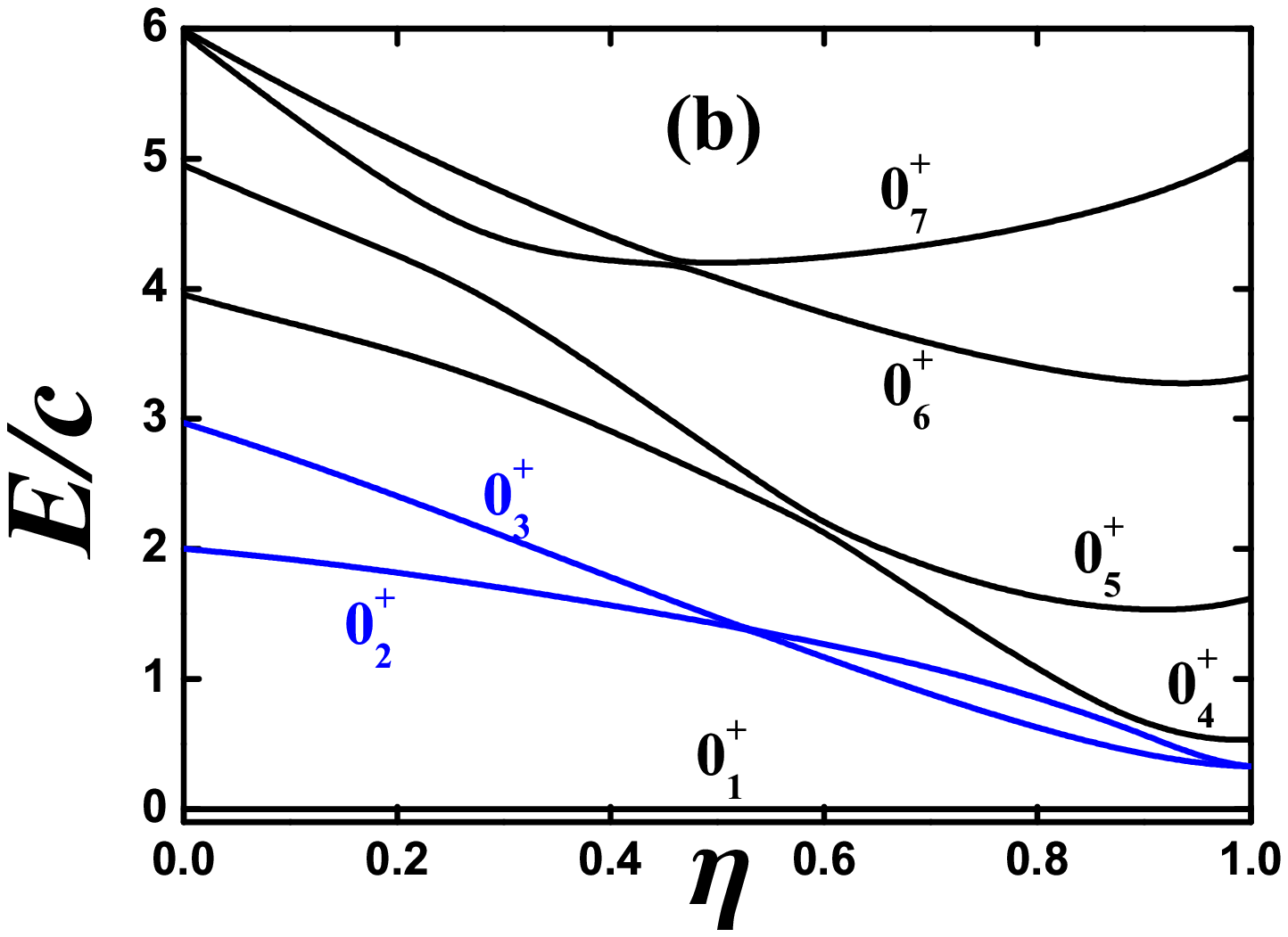}
\caption{(a) The evolutional behaviors of the $0^{+}$ states as a function of $\eta$ in $\hat{H}_{3}$ for $N=7$ from the U(5) symmetry limit to the SU(3) symmetry limit; (b) The evolutional behaviors of the $0^{+}$ states as a function of $\eta$ in $\hat{H}_{4}$ for $N=6$ from the U(5) symmetry limit to the SU(3) symmetry limit.}
\end{figure}

In this paper, we further investigate the new mechanism in Ref. \cite{Pan24}. This is very important. In \cite{Cheng25}, the new mechanism is found to be related to the SU(3) symmetry, so it greatly improve the understanding of the B(E2) anomaly, in a much broader perspective. Level-anticrossing plays an important role here, which occurs in the parameter region from the SU(3) symmetry limit to the O(6) symmetry limit, and is related to level-crossing phenomenon in the SU(3) symmetry limit. Level-crossing and level-anticrossing are two important phenomena in the IBM, see the analysis in \cite{Jolie02,Arias03}. In a previous paper \cite{Wang08}, One of the authors (T. Wang) and his coworkers found some interesting relationships, and in this paper, some deeper relationships are also found.

Based on this new relationship, we find that the two anomalous mechanisms in \cite{Wang20} and \cite{Pan24} can merge together. When the parameter varies from the SU(3) symmetry limit to the O(6) symmetry limit, a deep $B_{4/2}$ dip can appear. SU(3) symmetry and O(6) symmetry can have a deeper relationship. This new mechanism is related to the third interaction $[\hat{L}\times \hat{Q}_{\chi} \times \hat{L}]^{(0)}$ and the prolate-oblate asymmetric shape phase transition from the SU(3) symmetry limit to the O(6) symmetry limit.  These works will lead to a more complete explanatory framework for explaining the B(E2) anomaly. At last, the B(E2) anomaly in $^{170}$Os is discussed.

\section{Overview of level-anticrossing phenomenon}

In the IBM, the angular momentum $L$ is a conserved quantity. The evolutions of states with different angular momenta are not related to each other, and level-crossing can arise. Along the evolution from the U(5) symmetry limit to the O(6) symmetry limit, O(5) symmetry always exists, so states with the same angular momentum can have different O(5) quantum numbers $\tau$.
The evolutions of these states with different $\tau$ are also not related to each other, and level-crossing can occur. If we only discuss the evolutions between states with the same quantum number, level-crossing does not occur, but another effect can appear, which is level repulsion, or level-anticrossing. Level-crossing and level-anticrossing are two fundamental and important phenomena in the IBM, and even in the quantum mechanics.

Level-anticrossing is a universal phenomenon, which was discussed in \cite{Jolie02,Arias03} to understand the quantum phase transitions in the IBM. When the states have the same quantum numbers, these states can not be further distinguished, thus they may be related to each other. In some cases, two levels can get very close in energy, but because the repulsion between them also increases, they cannot crossover. This phenomenon is called level-anticrossing. In \cite{Jolie02,Arias03}, they mainly considered the quantum phase transition of the ground state, and discussed the relationship with level-anticrossing.

For a finite quantum system, the evolutions of low-energy excitations can be more complex and can also show interesting phenomena. To understand these, we should consider not only the ground state, but also the excited states. In nuclear structure, one excited state may have a new shape different from the ground state, or have a different rotational mode with different angular momentum. For the evolutions of $0^{+}$ states from the U(5) symmetry limit to the O(6) symmetry limit, the $0_{2}^{+}$ and $0_{3}^{+}$ states can crossover with each other, see Fig. 1(a). The Hamiltonian describing the evolutions is as follows
\begin{equation}
\hat{H}_{1}=c[(1-\eta)\hat{n}_{d}-\frac{\eta}{N}\hat{Q}_{0}\cdot \hat{Q}_{0}],
\end{equation}
here $\hat{n}_{d}$ is the $d$ boson number operator and $\hat{Q}_{0}$ is the O(6) quadrupole operator, $\eta$ and $c$ are two fitting parameters. At the U(5) symmetry side ($\eta=0$), the $0_{2}^{+}$ state has the O(5) quantum number $\tau=2$ and the $0_{3}^{+}$ state $\tau=3$. However at the O(6) symmetry side ($\eta=1$), they are just the opposite. Thus level-crossing arises. In \cite{Wang08}, this level-crossing was studied, which can be seen as a signature for the existence of the evolution from the U(5) symmetry limit to the O(6) symmetry limit.

A general Hamiltonian including the $\hat{H}_{1}$ is the $Q$-consistent formalism such as
\begin{equation}
\hat{H}_{2}=c[(1-\eta)\hat{n}_{d}-\frac{\eta}{N}\hat{Q}_{\chi}\cdot \hat{Q}_{\chi}],
\end{equation}
here $\hat{Q}_{\chi}$ is the general quadrupole operator. This Hamiltonian can describe various quadrupole shapes. If $\eta=0$, it can describe the spherical shape (the U(5) symmetry limit). If $\eta=1$ and $\chi=-\frac{\sqrt{7}}{2}$ ($\hat{Q}_{-\frac{\sqrt{7}}{2}}=\hat{Q}$), it can describe the prolate shape (the SU(3) symmetry limit). If $\eta=1$ and $\chi=0$, it can describe the $\gamma$-soft rotation (the O(6) symmetry limit). If $\eta=1$ and $\chi=\frac{\sqrt{7}}{2}$, it can describe the oblate shape (the $\overline{\textrm{SU(3)}}$ symmetry limit). Although this Hamiltonian is very simple, it is very valuable for understanding the collectivity of nuclei. When $\chi=-\frac{\sqrt{7}}{2}$, this Hamiltonian can be used to discuss the shape phase transition from the spherical shape to the prolate shape, which was studied in \cite{Pan03}, and the case with $N=10$ was shown. Clearly, in \cite{Pan03}, level-anticrossing do not appear. However, in \cite{Arias03}, the case with $N=50$ was studied, and the $0_{4}^{+}$ and $0_{5}^{+}$ states can get very close in energy at a specific $\eta$ value, and level-anticrossing can be observed. The same result can also arise when $N=15$, see Fig. 1(b). Thus the first mechanism for the emergence of level-anticrossing is to increase the boson number $N$.

When $\chi=0$, this is just the $\hat{H}_{1}$. This Hamiltonian can describe the shape phase transition from the spherical shape to the $\gamma$-soft rotation. $0_{2}^{+}$ and $0_{3}^{+}$ states can crossover with each other (Fig. 1(a)). In \cite{Wang08}, it was found that, if $\chi$ somewhat deviates from 0, level-crossing between the $0_{2}^{+}$, $0_{3}^{+}$ states becomes level-anticrossing due to the breaking of the O(5) symmetry and the complete discriminability of the two states. This provides the second mechanism for the emergence of level-anticrossing. If two states can be partially distinguished, level-anticrossing can occur.

In any case, two energy levels repel with each other because of the correlations between them, see \cite{Jolie02,Arias03}. In nuclei, level-crossing and level-anticrossing are not easy to distinguish due to the discrete change of the nucleon number $N$.

The SU3-IBM can present much complicated evolutional behaviors. In \cite{Wang22}, the spherical-like spectra was first proposed. The Hamiltonian is
\begin{equation}
\hat{H}_{3}=c[(1-\eta)\hat{n}_{d}+\eta(-\frac{\hat{C}_{2}[\textrm{SU(3)}]}{2N}+\kappa\frac{\hat{C}_{3}[\textrm{SU(3)}]}{2N^{2}})],
\end{equation}
here $\kappa$ is the parameter of the SU(3) third-order Casimir operator. When $\eta=1$, the Hamiltonian can describe the shape phase transition from the prolate shape to the oblate shape, which is different from the SU(3)-$\overline{\textrm{SU(3)}}$ description. $-\hat{C}_{2}[\textrm{SU(3)}]$ describes the prolate shape, and $\hat{C}_{3}[\textrm{SU(3)}]$ describe the oblate shape. For they both have the SU(3) symmetry, the shape transition can be discussed analytically, even for finite boson number $N$ \cite{Zhang12}. There exists a first-order phase transition point at $\kappa_{0}=\frac{3N}{2N+3}$. The $0^{+}$ states with SU(3) irrep. $(\lambda,\mu)$ satisfying $\lambda+2\mu=2N$ all crossover at this point, so it is also called SU(3) degenerate point. This is a simple example for level-crossing. It should be also noticed that, the energy levels evolve also in a linear way because the two quantities ($-\hat{C}_{2}[\textrm{SU(3)}]$ and $\hat{C}_{3}[\textrm{SU(3)}]$) commute with each other.

\begin{figure}[tbh]
\includegraphics[scale=0.33]{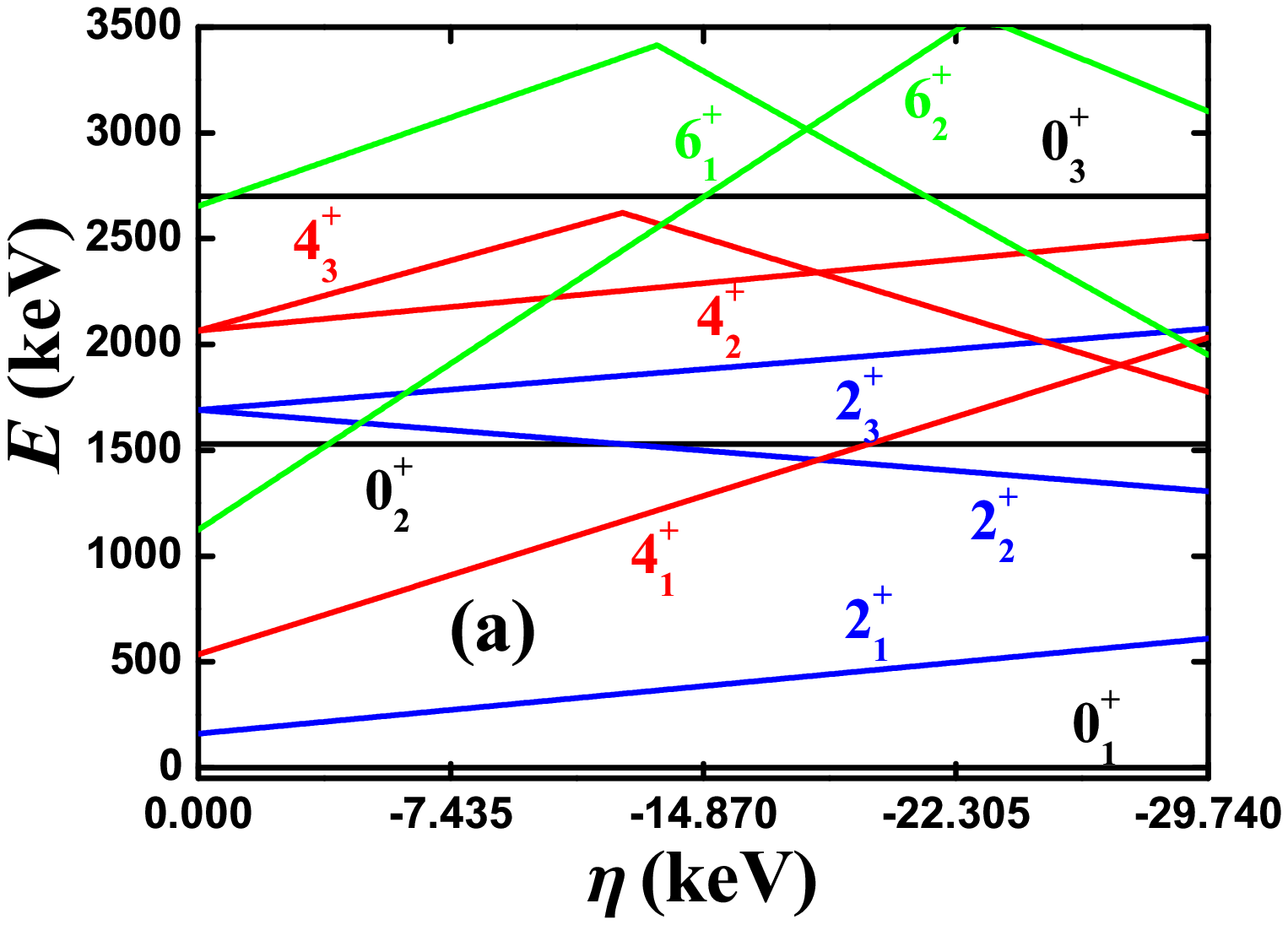}
\includegraphics[scale=0.33]{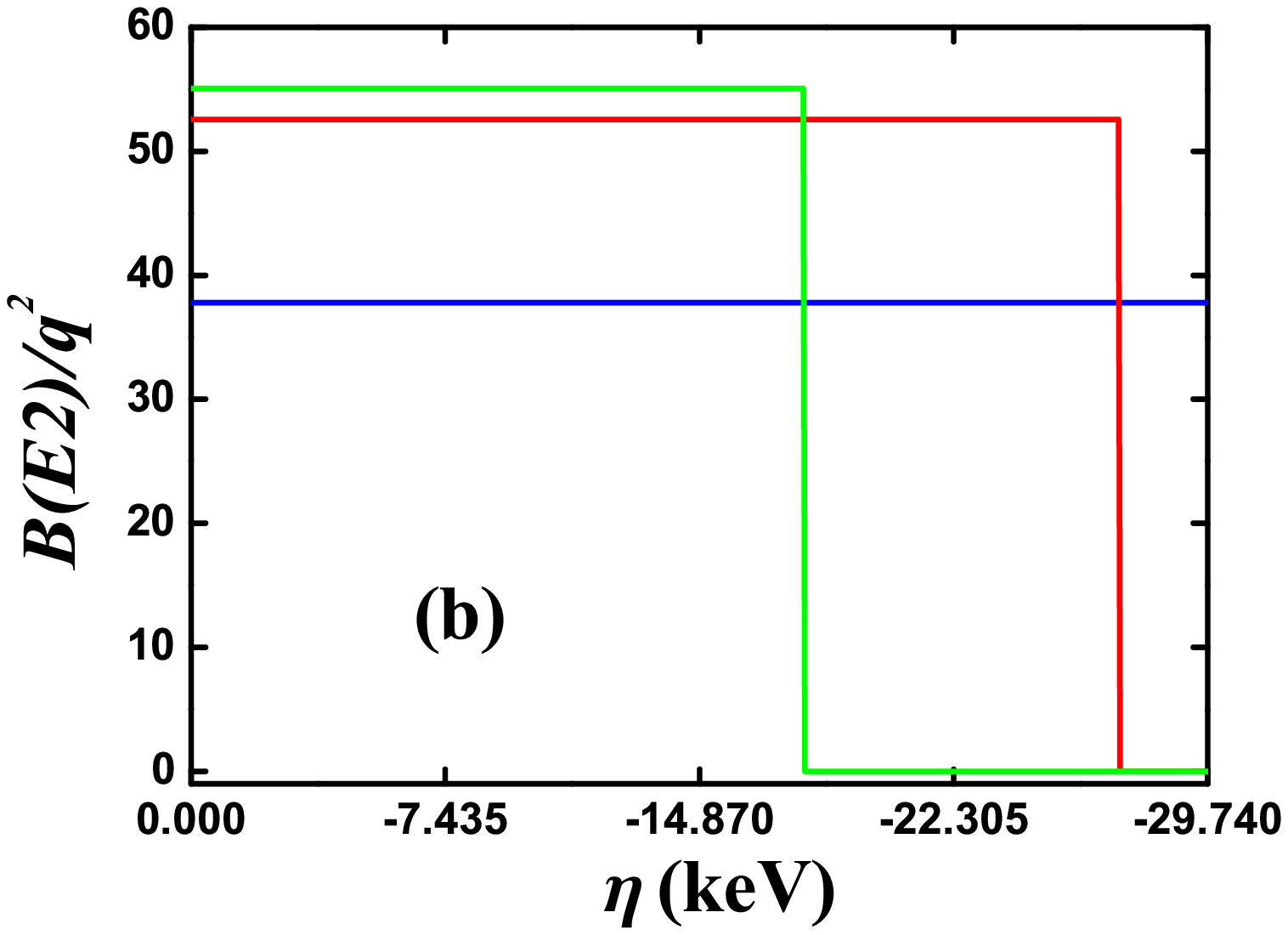}
\caption{(a) The evolutional behaviors of the partial low-lying levels as a function of $\eta$; (b) The evolutional behaviors of the $B(E2; 2_{1}^{+}\rightarrow 0_{1}^{+})$ (blue line) , $B(E2; 4_{1}^{+}\rightarrow 2_{1}^{+})$ (red line), $B(E2; 6_{1}^{+}\rightarrow 4_{1}^{+})$ (green line) as a function of $\eta$. The parameters are deduced from \cite{Pan24}.}
\end{figure}

If $\kappa_{0}=\frac{3N}{2N+3}$, the evolutional behaviors from the U(5) symmtry limit to the SU(3) degenerate point was discussed in \cite{Wang22}. The spherical-like spectra was first found at around $\eta=0.5$. Fig. 2(a) shows the evolutional behaviors of the low-lying $0^{+}$ states for $N=7$. Obviously, among the $0_{3}^{+}$, $0_{4}^{+}$, $0_{5}^{+}$, $0_{6}^{+}$ states, level-anticrossing can occur at various positions. At $\eta=0.5$, the energy of the $0_{3}^{+}$ state is nearly twice the one of the $0_{2}^{+}$ state, which results from level-anticrossing. The lowest five $0^{+}$ states has been confirmed in $^{106}$Pd \cite{WangPd}. Thus level-anticrossing plays a vital role in the SU3-IBM. Here the emergence of level-anticrossing results from addition of the SU(3) third-order interaction and the specific SU(3) degenerate point.

If the SU(3) fourth-order interactions is added, some new results can be obtained, which was discussed in \cite{Zhou23}. This Hamiltonian is as follows
\begin{equation}
\hat{H}_{4}=c[(1-\eta)\hat{n}_{d}+\eta(-\frac{\hat{C}_{2}[\textrm{SU(3)}]}{2N}+\zeta\frac{\hat{C}_{2}^{2}[\textrm{SU(3)}]}{2N^{3}})],
\end{equation}
here $\zeta$ is the parameter of the fourth-order interaction. When $\eta=1$, this Hamiltonian can be used to describe the shape phase transition from the prolate shape to any rigid triaxial deformations. This is an important finding. If rigid triaxial rotation really exists, the fourth-order interaction must be needed. Recently, Otsuka \emph{et al.} believed that, the large-deformed nuclei, such as $^{166}$Er, is in fact rigid triaxial shape with about $\gamma=8^{\circ}$ \cite{Otsuka19,Otsuka21,Otsuka}, and we really prove it within the SU3-IBM \cite{ZhouEr}.

When $\zeta=0.2232$, the SU(3) irrep of the ground state is (4,4), which presents a rigid triaxial shape with $\gamma=30^{\circ}$. Fig. 2(b) shows the evolutional behaviors of the low-lying $0^{+}$ states from the spherical shape to the rigid triaxial deformation. Prominent level-anticrossing of the $0_{2}^{+}$ and $0_{3}^{+}$ states arises, which seems to be similar to level-crossing of the $0_{2}^{+}$ and $0_{3}^{+}$ states from the U(5) symmetry limit to the O(6) symmetry limit in Fig. 1(a). Thus in the SU3-IBM, adding fourth-order interactions within the SU(3) symmetry limit can also induce the emergence of level-anticrossing.

In summary, there has been three mechanisms for the emergence of level-anticrossing in the IBM and the SU3-IBM: (1) increasing the boson number $N$, (2) the breaking of O(5) symmetry and (3) adding the SU(3) higher-order interactions within the SU(3) symmetry limit. When quantum states cannot be completely distinguished by the good quantum numbers, these states are associated, which is a common phenomenon in quantum mechanics. In the IBM or SU3-IBM, if not in the four dynamical symmetry limits, these quantum states are correlated to produce the level-anticrossing. Thus B(E2) anomaly and other accompanying anomalous phenomena can happen.

\section{Level-anticrossing in B(E2) anomaly}

B(E2) anomaly is a very unique phenomenon, which is only related to the $0_{1}^{+}$, $2_{1}^{+}$ and $4_{1}^{+}$  states in the yrast band. It is hard for nuclear structure theorists to imagine that it would exist if it were not experimentally determined that it does exist. After the discovery of the B(E2) anomaly in $^{168,170}$Os, $^{166}$W and $^{172}$Pt \cite{Grahn16,Saygi17,Cederwall18,Goasduff19}, it challenges all of the nuclear structure theories. The surprising thing is that there is still no explanation for any nuclear structure theory except for the SU3-IBM or the extended IBM theory.

\begin{figure}[tbh]
\includegraphics[scale=0.33]{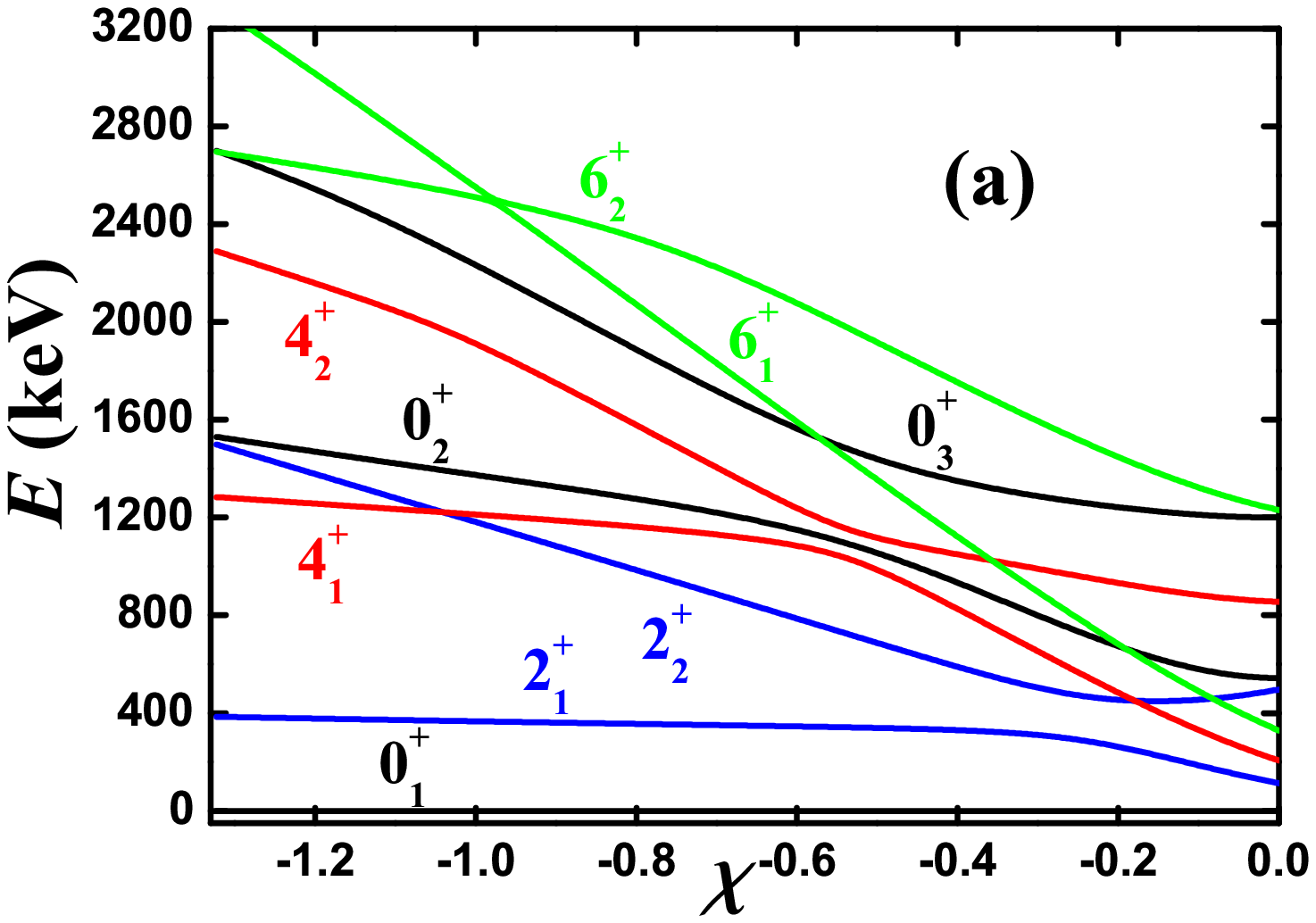}
\includegraphics[scale=0.33]{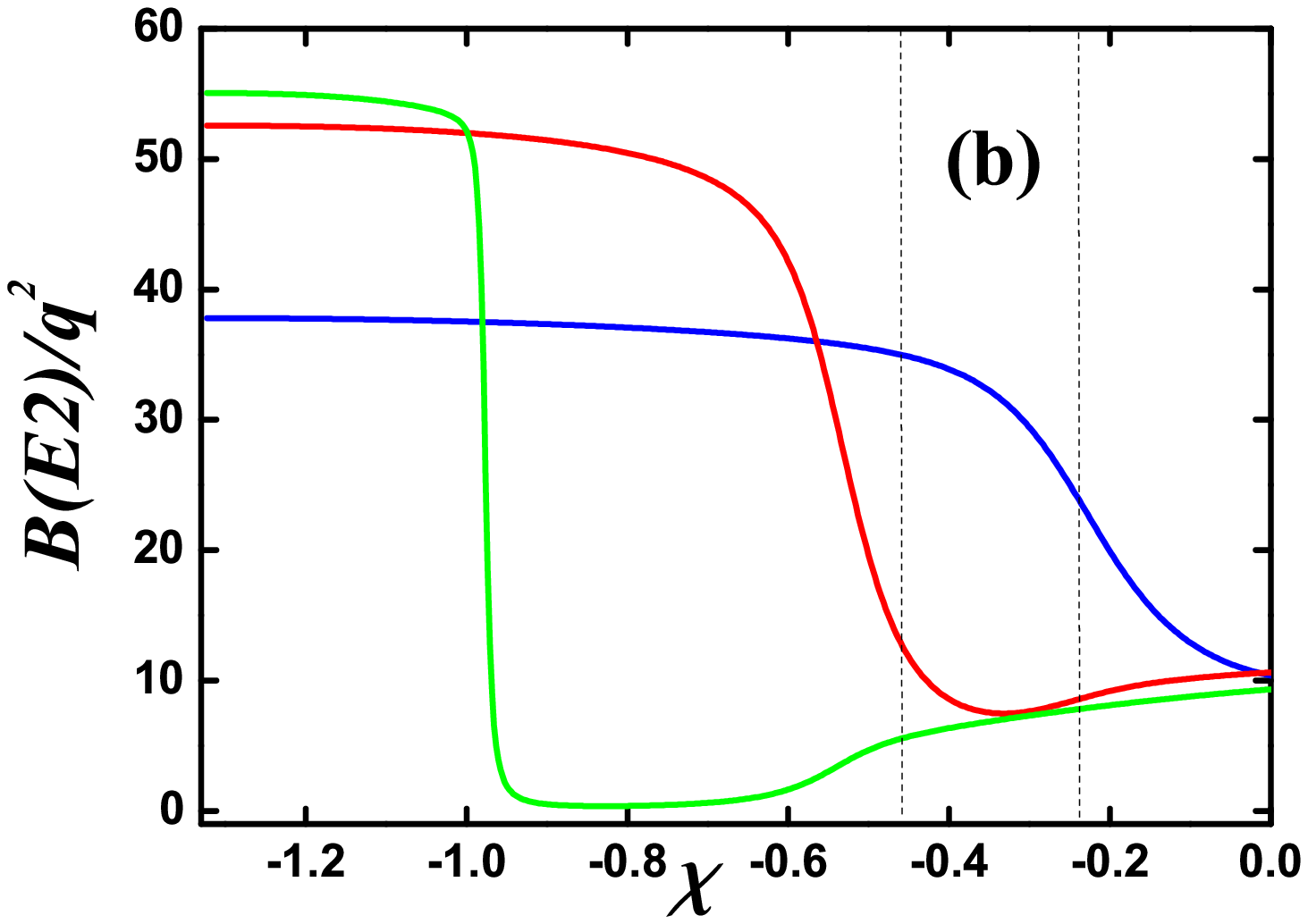}
\caption{(a) The evolutional behaviors of the partial low-lying levels as a function of $\chi$ for $N=9$ from the SU(3) symmetry limit to the O(6) symmetry limit; (b) The evolutional behaviors of the $B(E2; 2_{1}^{+}\rightarrow 0_{1}^{+})$ (blue line) , $B(E2; 4_{1}^{+}\rightarrow 2_{1}^{+})$ (red line), $B(E2; 6_{1}^{+}\rightarrow 4_{1}^{+})$ (green line) as a function of $\chi$. The parameters are deduced from \cite{Pan24}.}
\end{figure}

If the energy ratio $E_{4/2}\geq 2.0$, it is a signature for the collective excitations of nuclei. In general, the E2 transition ratio $B_{4/2}$ is larger than 1.0. In the B(E2) anomaly, this value can be smaller than 1.0, even reduces to 0.33. If $0_{1}^{+}$, $2_{1}^{+}$, $4_{1}^{+}$ states belong to the same rotational band, it really seems to be very strange. However, in the rigid triaxial rotor, this phenomenon really can occur, which was first discovered by Y. Zhang \emph{et al.} \cite{zhang14}. In this finding, when angular momentum $L$ of the state in the yrast band increases, the E2 transitional rate to the $L-2$ state can decrease, but varies in a way from slow to fast. The $B_{4/2}$ seems to be larger than 0.5, can not reduce to 0.33. This needs to be discussed in future.

\begin{figure}[tbh]
\includegraphics[scale=0.33]{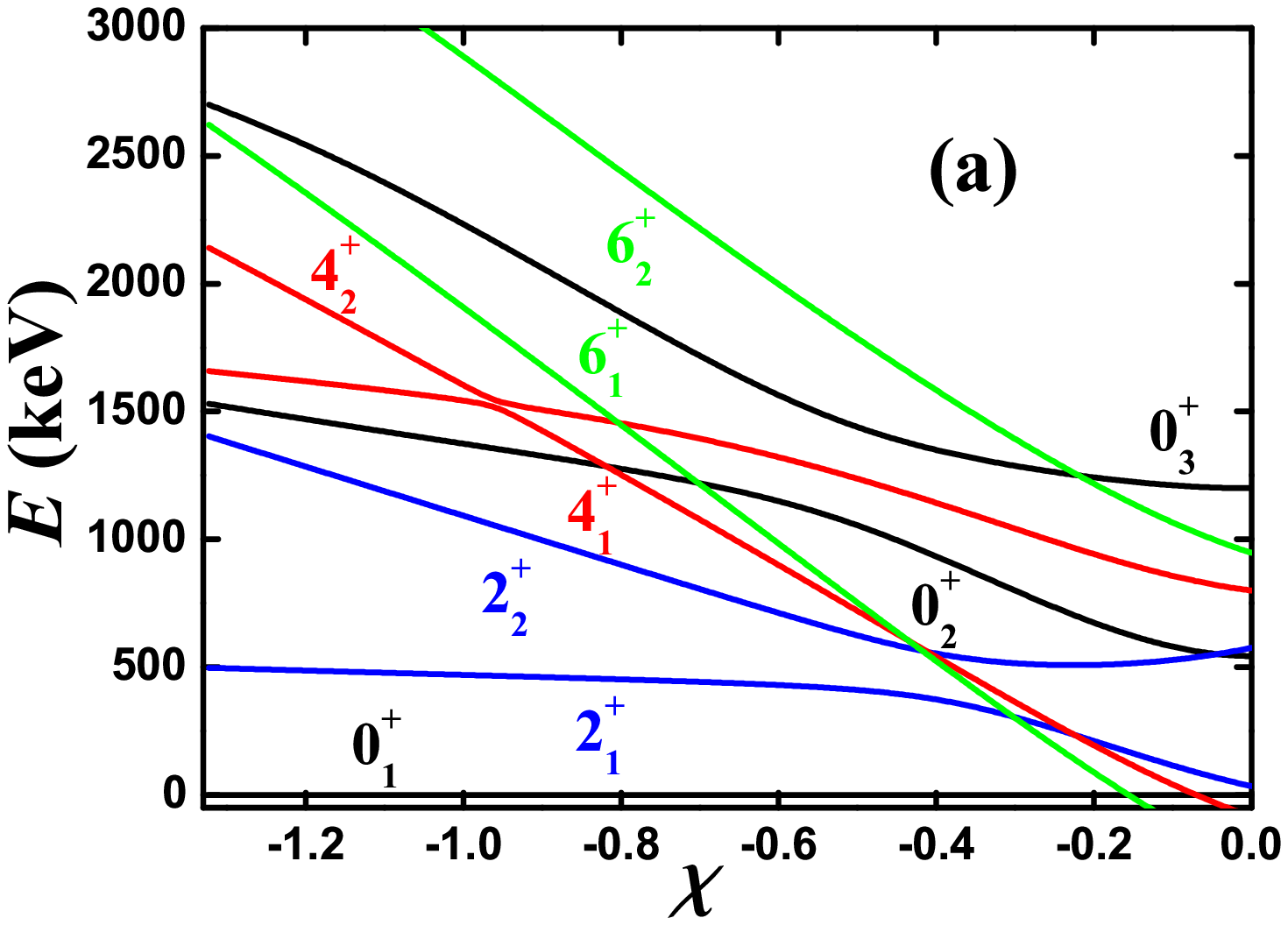}
\includegraphics[scale=0.33]{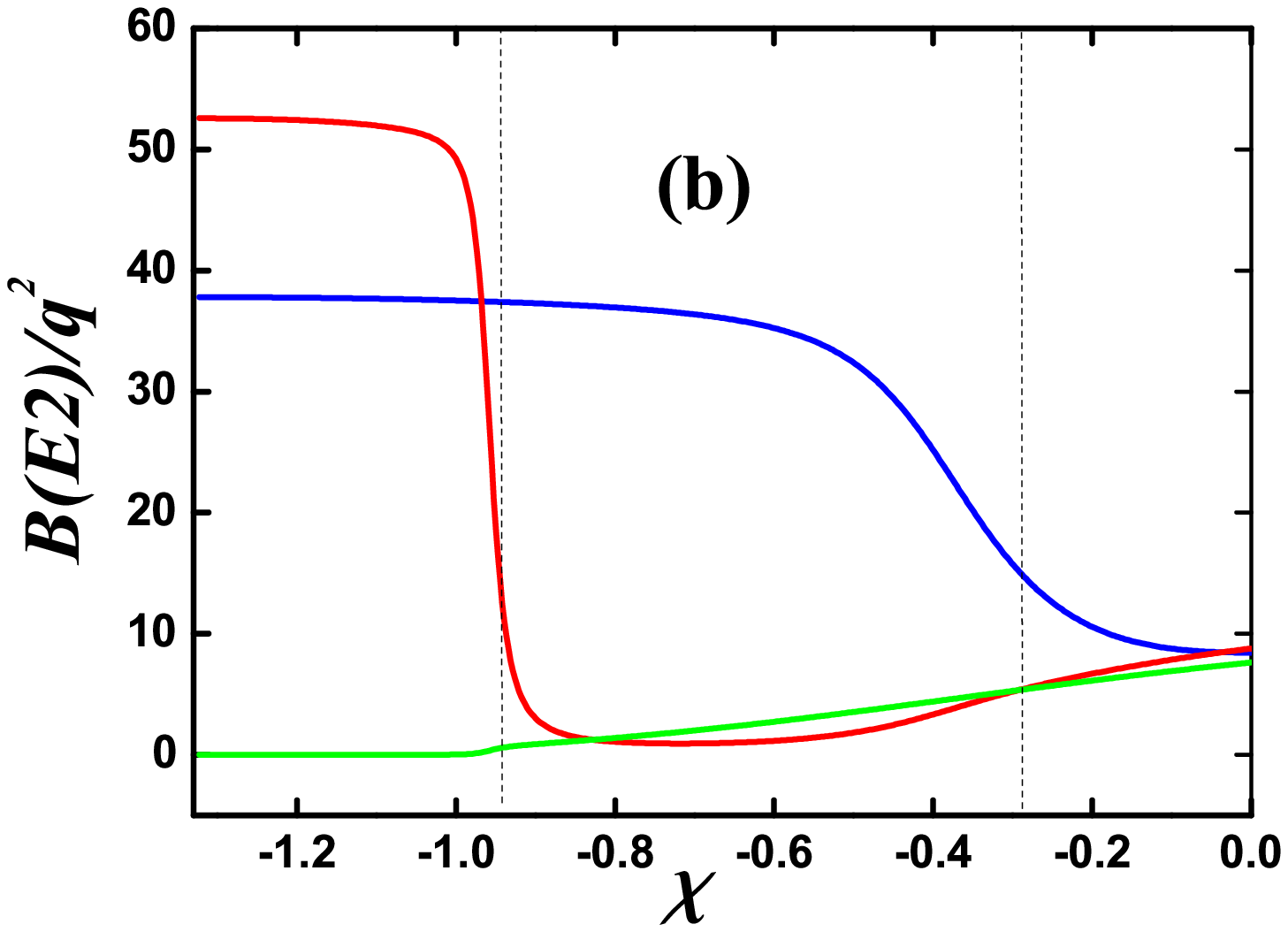}
\caption{(a) The evolutional behaviors of the partial low-lying levels as a function of $\chi$ for $N=9$ from the SU(3) symmetry limit to the O(6) symmetry limit; (b) The evolutional behaviors of the $B(E2; 2_{1}^{+}\rightarrow 0_{1}^{+})$ (blue line) , $B(E2; 4_{1}^{+}\rightarrow 2_{1}^{+})$ (red line), $B(E2; 6_{1}^{+}\rightarrow 4_{1}^{+})$ (green line) as a function of $\chi$. The parameters are deduced from \cite{Pan24}.}
\end{figure}

One of the most direct reasons for the emergence of the B(E2) anomaly is that in the SU(3) symmetry limit, these three states $0_{1}^{+}$, $2_{1}^{+}$ and $4_{1}^{+}$ do not actually belong to the same rotational band. In \cite{Wang20}, one of the authors (T. Wang) first provided such an explanation. In the SU(3) limit, the SU(3) third-order interaction $[\hat{L}\times \hat{Q} \times \hat{L}]^{(0)}$ can induce level-crossing between the $4_{1}^{+}$ state and one other $4^{+}$ state. It should be noticed that this level-crossing is not easily realized by other higher-order interactions. In \cite{Wangtao}, one of the authors (T. Wang) showed that, in the O(6) limit, this similar result can not be obtained. This gives a strong limit for the explanation of the B(E2) anomaly if up to fourth-order interactions are considered.

\begin{figure}[tbh]
\includegraphics[scale=0.33]{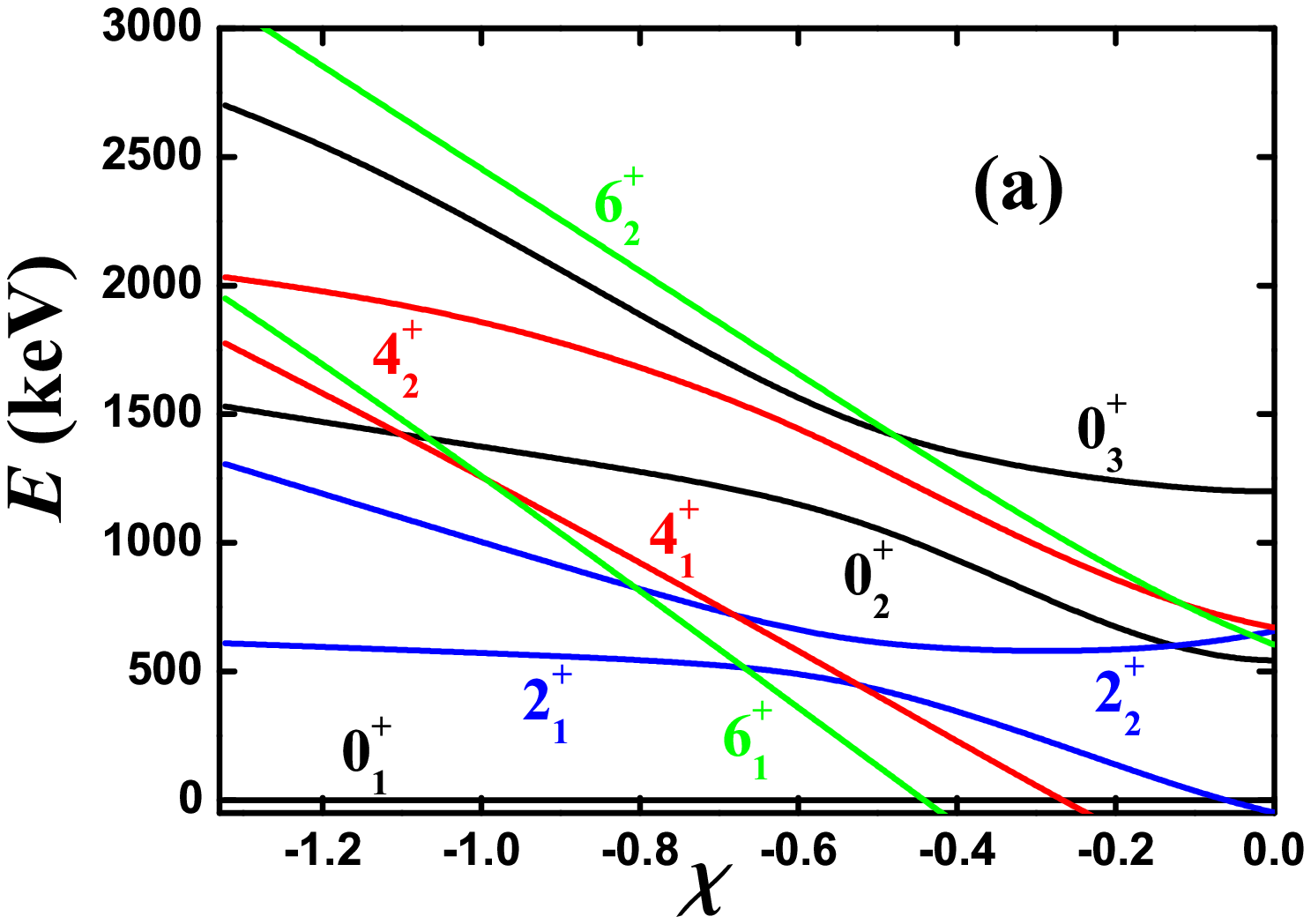}
\includegraphics[scale=0.33]{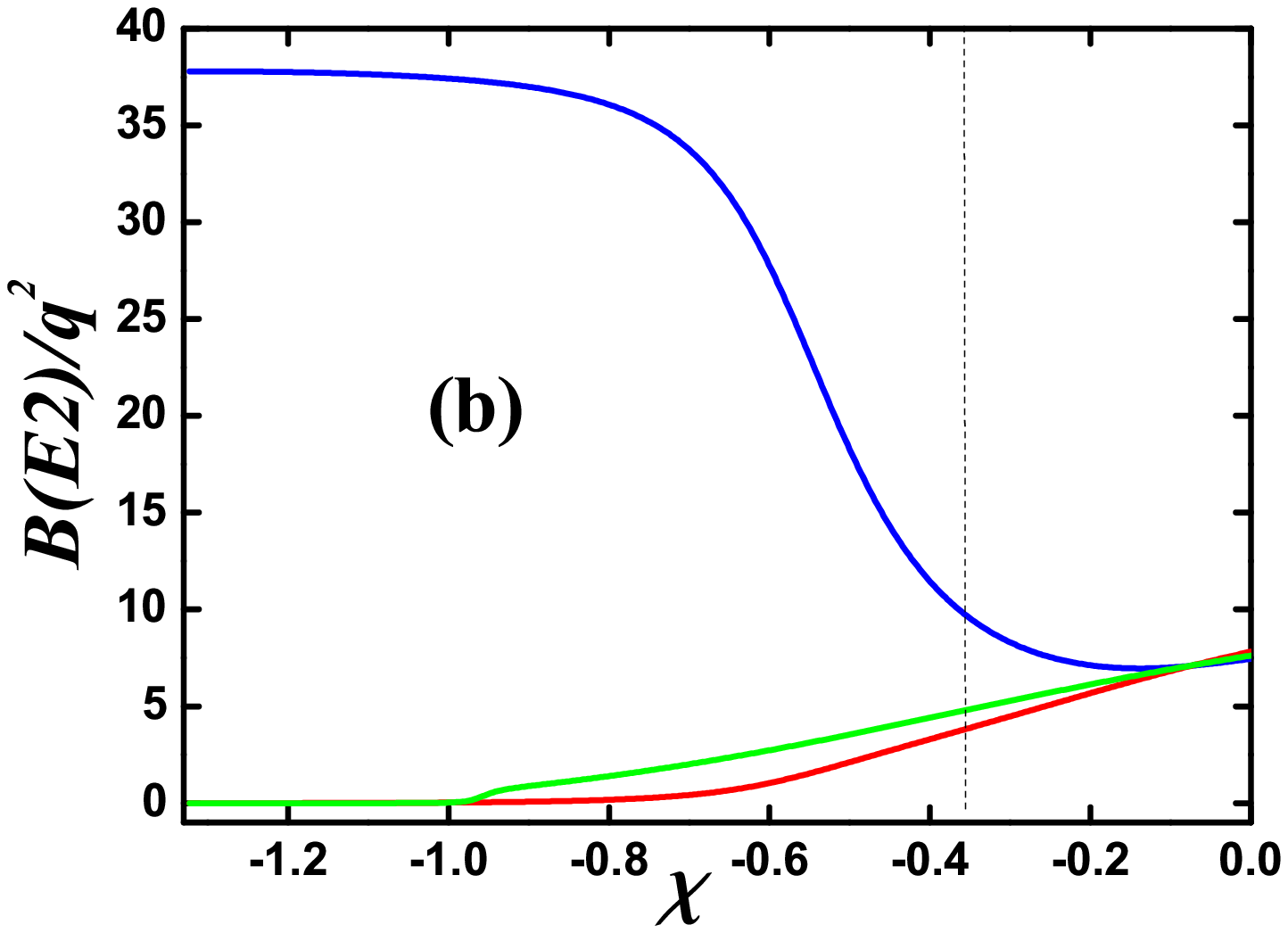}
\caption{(a) The evolutional behaviors of the partial low-lying levels as a function of $\chi$ for $N=9$ from the SU(3) symmetry limit to the O(6) symmetry limit; (b) The evolutional behaviors of the $B(E2; 2_{1}^{+}\rightarrow 0_{1}^{+})$ (blue line) , $B(E2; 4_{1}^{+}\rightarrow 2_{1}^{+})$ (red line), $B(E2; 6_{1}^{+}\rightarrow 4_{1}^{+})$ (green line) as a function of $\chi$. The parameters are deduced from \cite{Pan24}.}
\end{figure}

Recently, the B(E2) anomaly has been further studied. To find out whether these new mechanisms are related to the SU(3) symmetry, and to better distinguish between these different mechanisms, in the previous paper, the concept ``SU(3) analysis" was proposed \cite{Cheng25}. If a Hamiltonian has the SU(3) limit, we should first discuss whether the B(E2) anomaly exists in the SU(3) symmetry limit. When the parameter of the SU(3) third-order interaction $[\hat{L}\times \hat{Q} \times \hat{L}]^{(0)}$ varies from 0, the evolutional behaviors of the low-lying levels and the E2 transitional rates between these levels are studied. If level-crossing between the $4_{1}^{+}$ state and one other $4^{+}$ state can occur, $B_{4/2}=0$ and the B(E2) anomaly exists.

In a recent paper by F. Pan \emph{et al.}, a new mechanism was found for explaining the B(E2) anomaly \cite{Pan24}. In their study, the O(6) symmetry limit is also included. The B(E2) anomaly was found at the parameter region from the SU(3) symmetry limit to the O(6) symmetry limit. This seems very different from previous explanations. They used a Hamiltonian as follows
\begin{equation}
\hat{H}_{5}=\varepsilon_{d}\hat{n}_{d}-\kappa \hat{Q}_{\chi}\cdot \hat{Q}_{\chi}+\eta[\hat{L}\times \hat{Q}_{\chi} \times \hat{L}]^{(0)}+f\hat{L}^{2}.
\end{equation}
Compared to the $\hat{H}_{2}$, the $[\hat{L}\times \hat{Q}_{\chi} \times \hat{L}]^{(0)}$ interaction is added. In $\hat{H}_{2}$, the B(E2) anomaly can not appear, so the third-order interaction is important for the emergence of the B(E2) anomaly. In the SU(3) symmetry limit, this interaction can generate level-crossing and the B(E2) anomaly, but in the O(6) symmetry limit, this can not occur \cite{Wangtao}. These results imply that the SU(3) symmetry is vital for the B(E2) anomaly. In \cite{Pan24}, it seems to be not so. In the previous paper, the case in \cite{Pan24} was also discussed with the SU(3) analysis using the example of $^{168}$Os \cite{Cheng25}. Fig. 3 shows the results of $^{170}$Os with the parameters in \cite{Pan24} when $\chi=-\frac{\sqrt{7}}{2}$ and $\eta$ varies from 0 to -29.74 keV ($\varepsilon_{d}=0.0$ for the SU(3) analysis and $\kappa=30.0$ keV, $f=15.5$ keV), and the middle point is the case of the SU(3) analysis in \cite{Pan24}. Clearly, the $4_{1}^{+}$ state and one other $4^{+}$ state intersect at $\eta=-27.15$ keV and the $6_{1}^{+}$ state and one other $6^{+}$ state intersect at $\eta=-17.9$ keV (see Fig. 3(a)).

For understanding the B(E2) anomaly, the $B(E2)$ values are necessary. The $E2$ operator is defined as
\begin{equation}
\hat{T}(E2)=q\hat{Q}_{\chi},
\end{equation}
where $q$ is the boson effective charge. In Fig. 3(b) the E2 transitional rates of $B(E2;2_{1}^{+}\rightarrow 0_{1}^{+})$, $B(E2;4_{1}^{+}\rightarrow 2_{1}^{+})$ and $B(E2;6_{1}^{+}\rightarrow 4_{1}^{+})$ are shown, and clearly the B(E2) anomaly really exists. The value of $B(E2;4_{1}^{+}\rightarrow 2_{1}^{+})$ is 0 when $\eta\leq -27.15$ keV. At the middle point, the B(E2) anomaly can not occur. Thus the new mechanism is related to the SU(3) symmetry, but in a \emph{hidden} way. This is very interesting. This implies that previous understanding with $B(E2;4_{1}^{+}\rightarrow 2_{1}^{+})=0$ is inefficient.

Now we discuss the emergence of the B(E2) anomaly in \cite{Pan24}. Fig. 4(a) shows the evolutional behaviors of the low-lying levels when the parameter $\chi$ changes from $-\frac{\sqrt{7}}{2}$ (the SU(3) symmetry limit) to 0 (the O(6) symmetry limit). The parameters are deduced from \cite{Pan24} ($\kappa=30.0$ keV, $\eta=-14.87$ keV and $f=15.5$ keV). In \cite{Pan24} $\varepsilon_{d}=60.0$ keV, but here $\varepsilon_{d}=0$ keV. Thus the middle point ($\eta=-14.87$ keV) in Fig. 3(a) is the SU(3) symmetry limit of Fig. 4(a). Prominent level-anticrossing can be seen for the $4_{1}^{+}$, $4_{2}^{+}$ states and $6_{1}^{+}$, $6_{2}^{+}$ states. In Fig. 4(b), at the SU(3) symmetry limit and the O(6) symmetry limit, the $B_{4/2}$ values are larger than 1.0. Around $\chi=-0.35$, the B(E2) anomaly does arise. The parameter region between the two dashed line is for $B_{4/2}\leq 0.38$. 0.38 is the $B_{4/2}$ value in $^{170}$Os. Thus an anomalous $B_{4/2}$ dip exists. The $B(E2;6_{1}^{+}\rightarrow 4_{1}^{+})$ anomaly also exists. Thus the new mechanism in \cite{Pan24} is related to level-anticrossing.

\begin{figure}[tbh]
\includegraphics[scale=0.33]{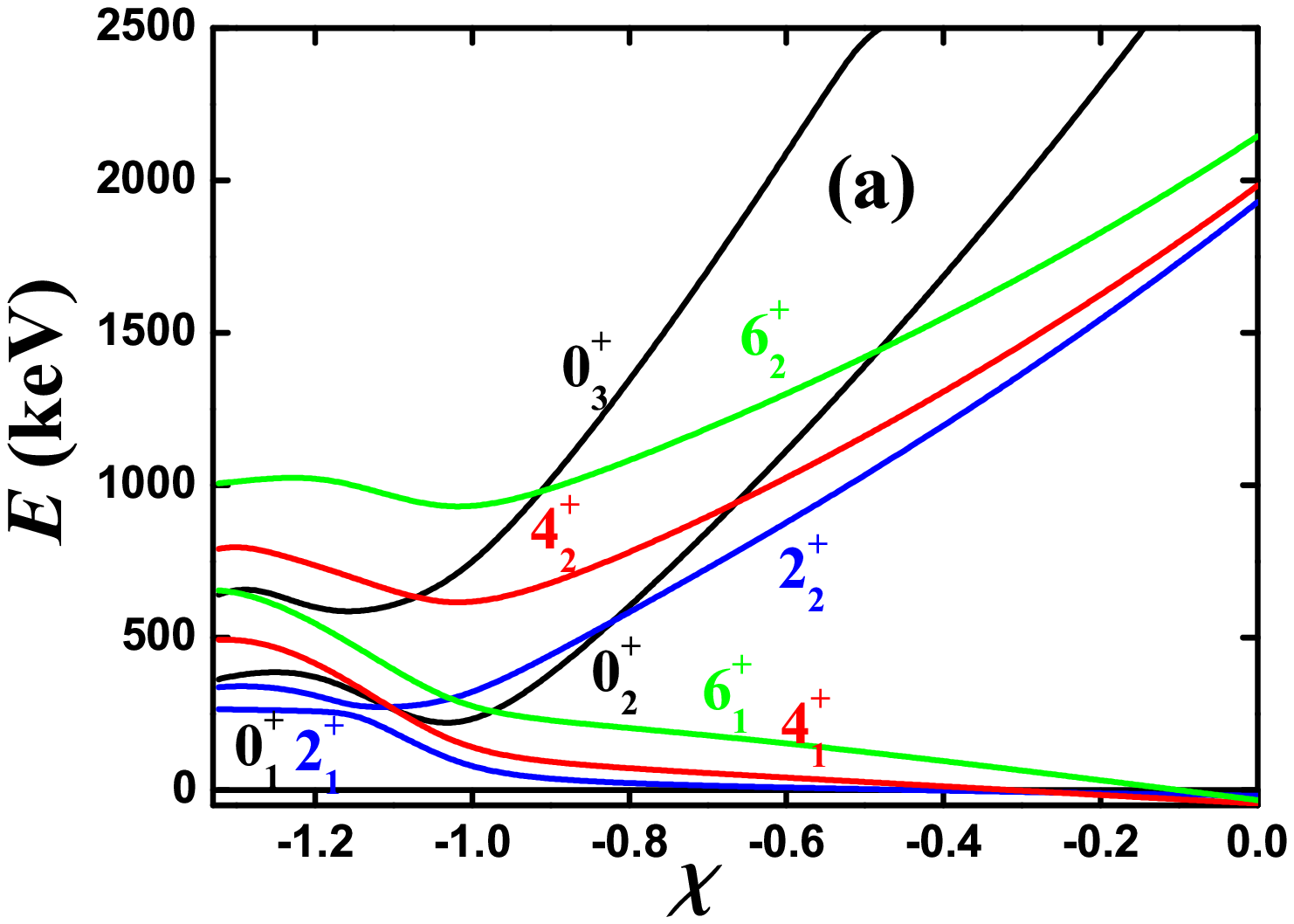}
\includegraphics[scale=0.33]{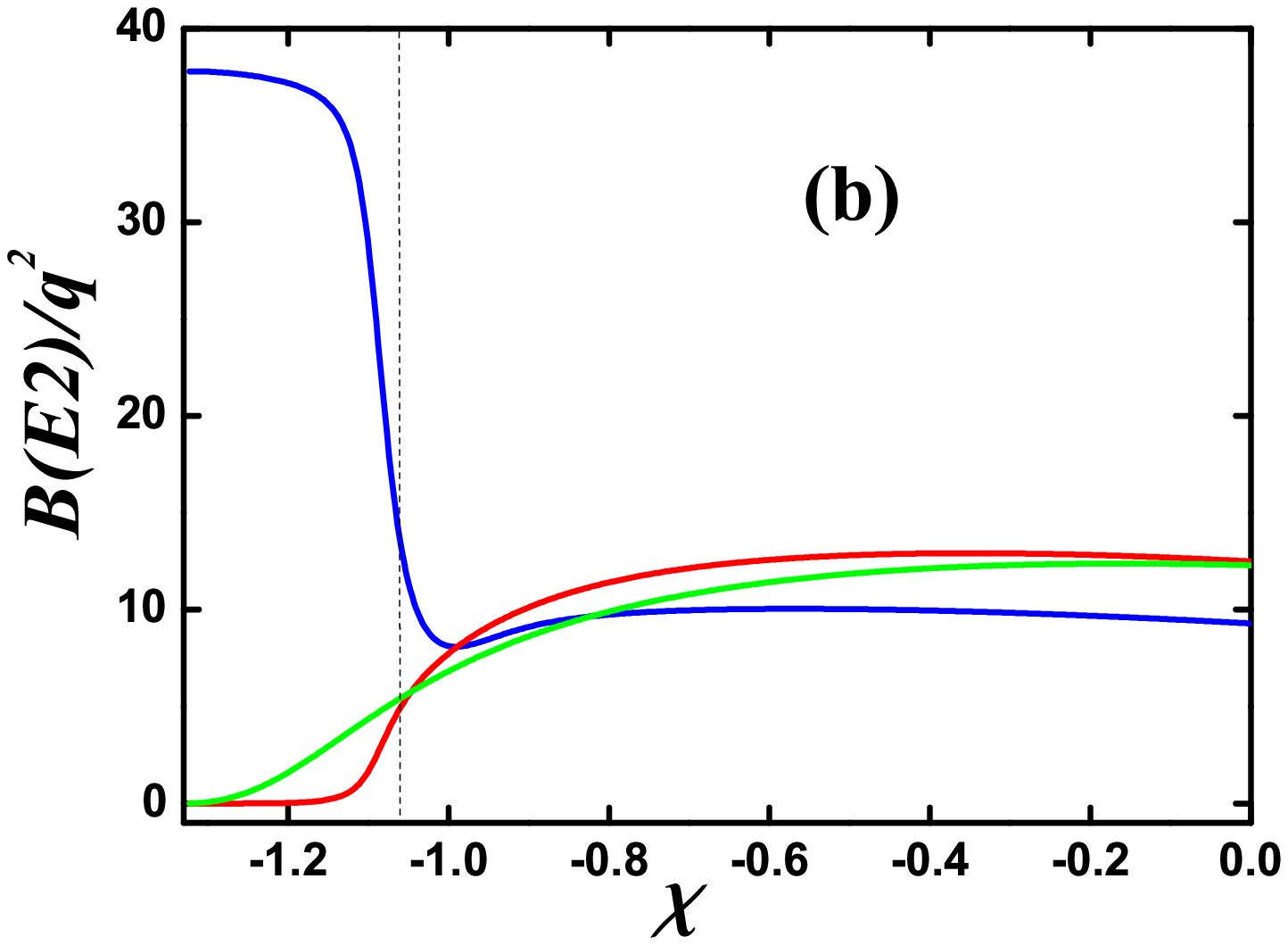}
\caption{(a) The evolutional behaviors of the partial low-lying levels as a function of $\eta$; (b) The evolutional behaviors of the $B(E2; 2_{1}^{+}\rightarrow 0_{1}^{+})$ (blue line) , $B(E2; 4_{1}^{+}\rightarrow 2_{1}^{+})$ (red line), $B(E2; 6_{1}^{+}\rightarrow 4_{1}^{+})$ (green line) as a function of $\eta$. The parameters are deduced from \cite{Wang20}.}
\end{figure}

In Fig. 4, $\eta=-14.87$ keV. Now we change $\eta$ to study level-anticrossing in detail. Fig. 5(a) shows the evolutional behaviors of the low-lying levels when the parameter $\chi$ changes from $-\frac{\sqrt{7}}{2}$  to 0 and $\eta=-22.305$ keV. Other parameters are the same as the ones in Fig. 4. In Fig. 3(a), from $-14.87$ keV to $-22.307$ keV, the $6_{1}^{+}$ state crossovers with one other $6^{+}$ state. In Fig. 5(a), level-anticrossing of the two $6^{+}$ states disappears. Level-anticrossing of the $4_{1}^{+}$ and $4_{2}^{+}$ states becomes more prominent and moves to the SU(3) symmetry side. In Fig. 5(b), the B(E2) anomaly seems to be more prominent too, and exists in a broader parameter region. The parameter region between the two dashed line for $B_{4/2}\leq0.38$ becomes larger. However, in the most part of this anomalous region, the $B(E2;6_{1}^{+}\rightarrow 4_{1}^{+})$ value is nearly the same as the one of $B(E2;4_{1}^{+}\rightarrow 2_{1}^{+})$, which is an important signature for this new mechanism. However, at the position of the left dashed line, the $B(E2;6_{1}^{+}\rightarrow 4_{1}^{+})$ value is much smaller than the one of $B(E2;4_{1}^{+}\rightarrow 2_{1}^{+})$.

\begin{figure}[tbh]
\includegraphics[scale=0.33]{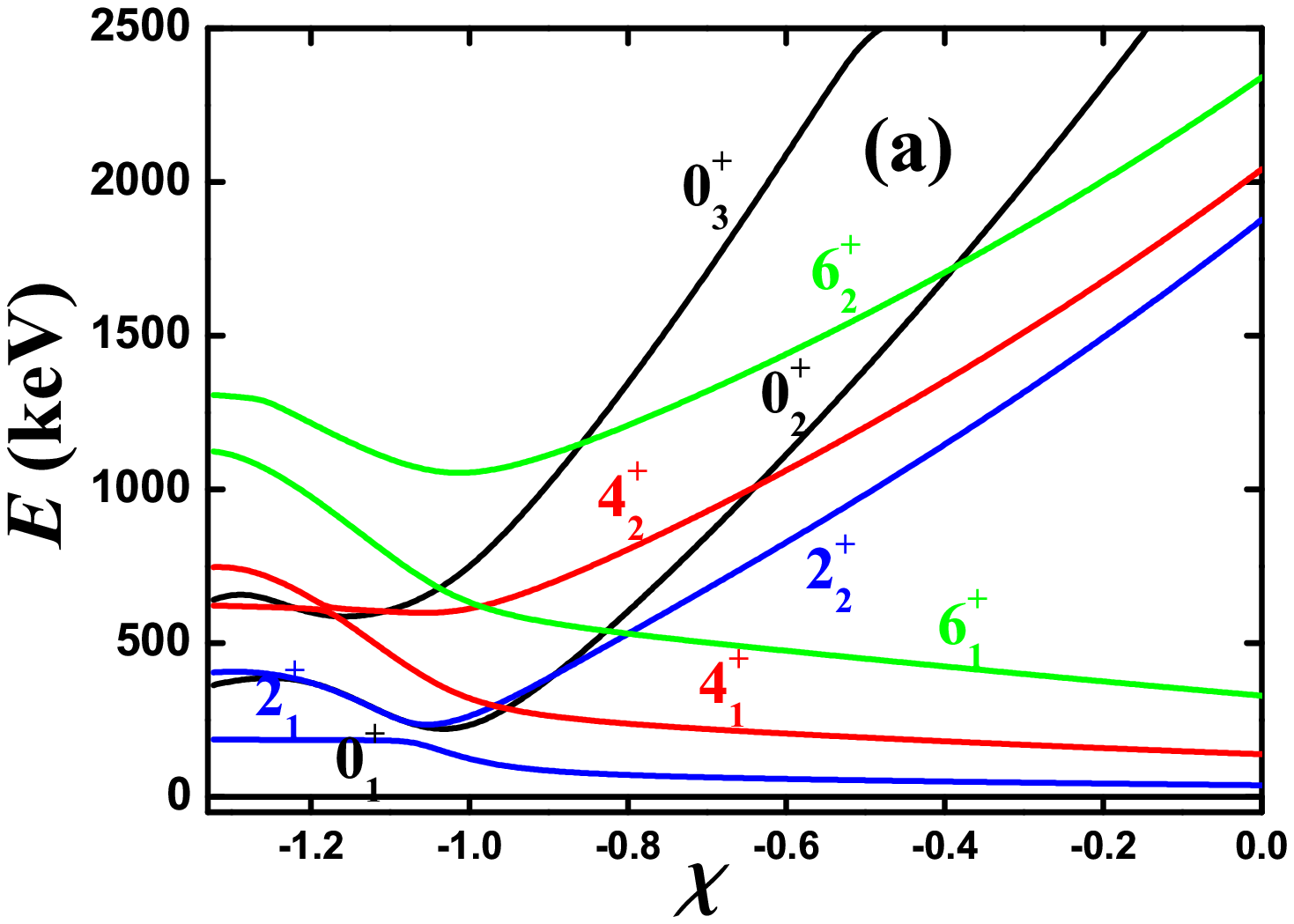}
\includegraphics[scale=0.33]{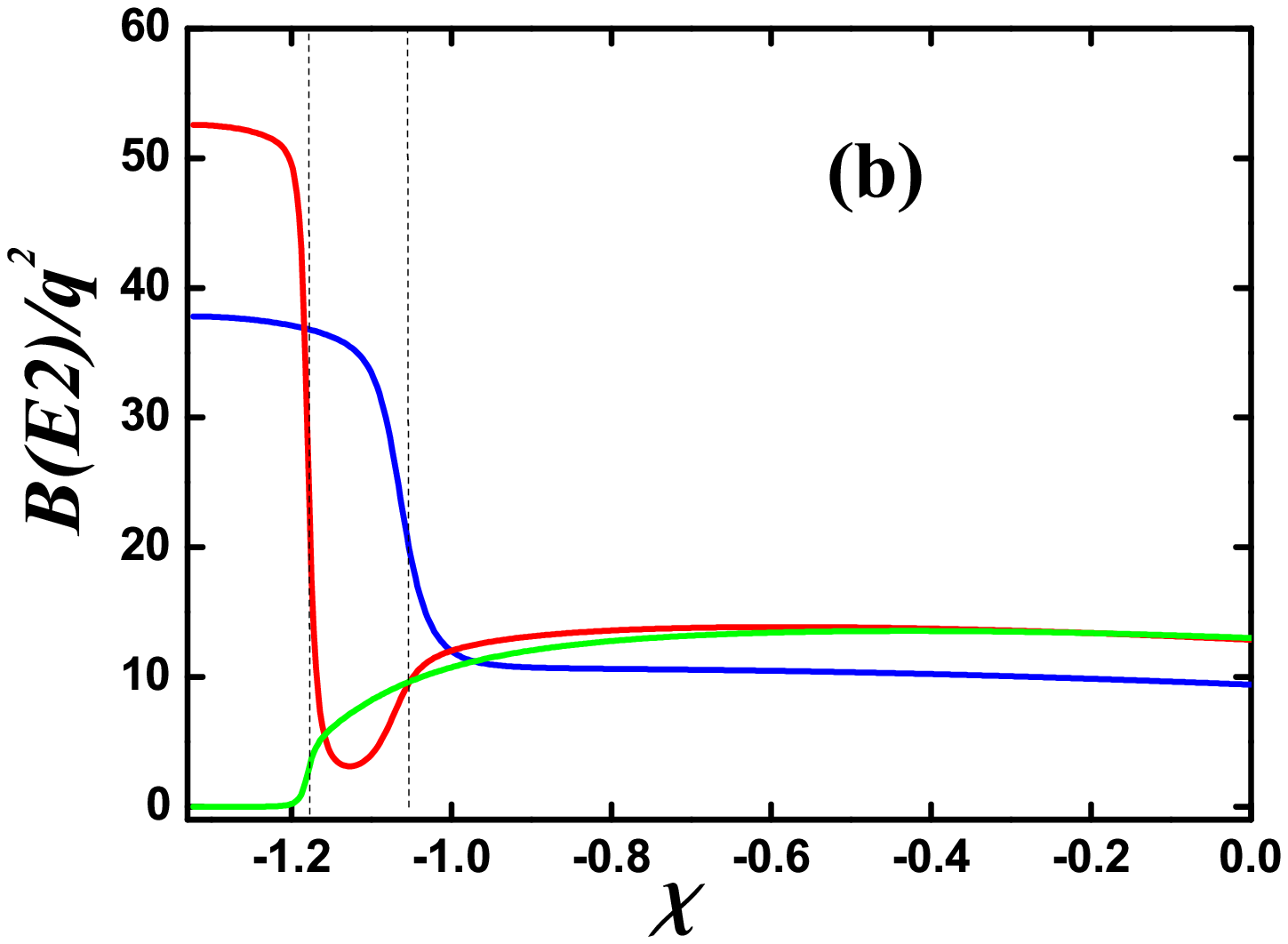}
\caption{(a) The evolutional behaviors of the partial low-lying levels as a function of $\eta$; (b) The evolutional behaviors of the $B(E2; 2_{1}^{+}\rightarrow 0_{1}^{+})$ (blue line) , $B(E2; 4_{1}^{+}\rightarrow 2_{1}^{+})$ (red line), $B(E2; 6_{1}^{+}\rightarrow 4_{1}^{+})$ (green line) as a function of $\eta$. The parameters are deduced from \cite{Wang20}.}
\end{figure}

In Fig. 6, $\eta$ becomes to $-29.74$ keV. In Fig. 3(a), from $-22.307$ keV to $-29.74$ keV, the $4_{1}^{+}$ state intersects with one other $4^{+}$ state. Fig. 6(a) shows the evolutional behaviors of the low-lying levels when the parameter $\chi$ changes from $-\frac{\sqrt{7}}{2}$ to 0. Other parameters are the same as the ones in Fig. 4. Clearly, level-anticrossing of the $4_{1}^{+}$ and $4_{2}^{+}$ states also disappears. In Fig. 6(b) the B(E2) anomaly occurs from the SU(3) symmetry limit and exists in a more broader parameter region. The left side of the dashed line is for $B_{4/2}\leq 0.38$. These discussions reveal that, \emph{level-crossing in the SU(3) symmetry limit is related to level-anticrossing in the parameter region from the SU(3) symmetry limit to the O(6) symmetry limit}. This is the most important finding in this paper. Thus this new mechanism is really related to the SU(3) symmetry

\begin{figure}[tbh]
\includegraphics[scale=0.33]{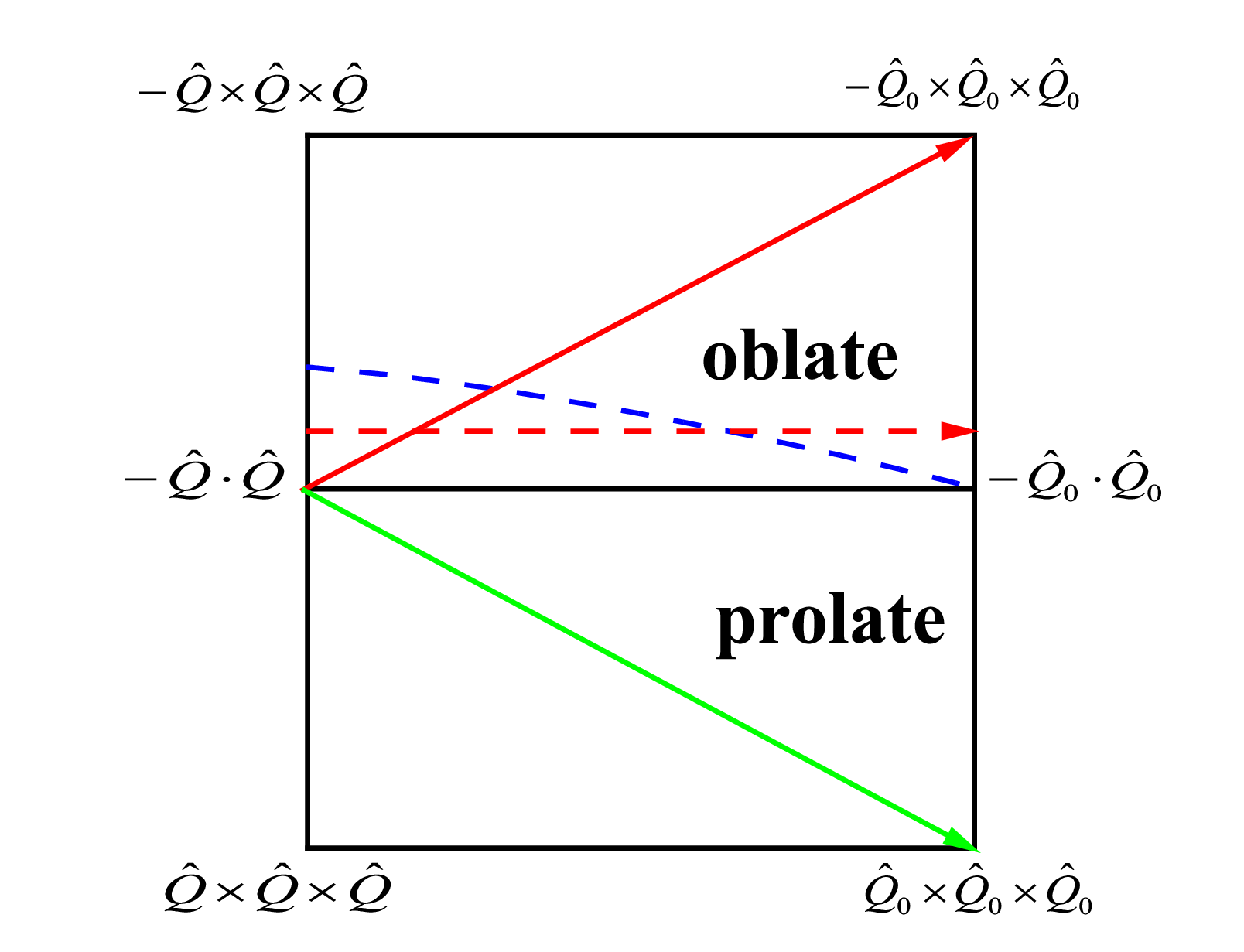}
\caption{The quadrupole deformation part of the phase diagram in the extended model proposed by Fortunato et al. [9]. The upper part above the dashed blue line represents the oblate shape while the lower part represents the prolate shape. The green line presents the evolution in $\hat{H}_{8}$ while the red line presents the one in $\hat{H}_{9}$. }
\end{figure}

This finding is very interesting. In previous IBM, from $-\hat{Q}\cdot \hat{Q}$ (the SU(3) symmetry limit) to $-\hat{Q}_{0}\cdot \hat{Q}_{0}$ (the O(6) symmetry limit), there is no level-crossing or level-anticrossing phenomenon. In $\hat{H}_{5}$, when the third-order interaction $[\hat{L}\times \hat{Q}_{\chi} \times \hat{L}]^{(0)}$ is introduced, level-crossing from $-\hat{Q}\cdot \hat{Q}$ to $-\hat{Q}\cdot \hat{Q}+\frac{\eta}{\kappa}[\hat{L}\times \hat{Q} \times \hat{L}]^{(0)}$ in the SU(3) symmetry limit or level-anticrossing from $-\hat{Q}\cdot \hat{Q}+\frac{\eta}{\kappa}[\hat{L}\times \hat{Q} \times \hat{L}]^{(0)}$ to $-\hat{Q}_{0}\cdot \hat{Q}_{0}+\frac{\eta}{\kappa}[\hat{L}\times \hat{Q}_{0} \times \hat{L}]^{(0)}$ must occur, and only one result can appear. In \cite{Wangtao}, in the O(6) symmetry limit, it was found that, when the negative $[\hat{L}\times \hat{Q}_{0} \times \hat{L}]^{(0)}$ interaction is added to the $-\hat{Q}_{0}\cdot \hat{Q}_{0}$, it presents an oblate shape. Thus from $-\hat{Q}\cdot \hat{Q}$ to $-\hat{Q}\cdot \hat{Q}+\frac{\eta}{\kappa}[\hat{L}\times \hat{Q} \times \hat{L}]^{(0)}$, and then to $-\hat{Q}_{0}\cdot \hat{Q}_{0}+\frac{\eta}{\kappa}[\hat{L}\times \hat{Q}_{0} \times \hat{L}]^{(0)}$ presents the prolate-oblate asymmetric shape phase transition.

\section{A general explanatory framework for the B(E2) anomaly}

Now we establish a general explanatory framework for the B(E2) anomaly with up to third-order interactions. Fourth-order interactions are considered in future because their discussions are much complicated. Thus the Hamiltonian is as follows
\begin{eqnarray}
\hat{H}_{7}&=&\varepsilon_{d}\hat{n}_{d}-\kappa \hat{Q}_{\chi}\cdot \hat{Q}_{\chi}+\zeta[\hat{Q}_{\chi} \times \hat{Q}_{\chi} \times \hat{Q}_{\chi}]^{(0)}  \nonumber\\
&&+\eta[\hat{L}\times \hat{Q}_{\chi} \times \hat{L}]^{(0)}+f\hat{L}^{2}.
\end{eqnarray}
If $\chi=-\frac{\sqrt{7}}{2}$, this Hamiltonian has been discussed for describing the B(E2) anomaly in \cite{Wang20}. Here the $\chi$ value can vary from $-\frac{\sqrt{7}}{2}$ to 0. This Hamiltonian combines the ideas in \cite{Wang20} and \cite{Pan24}.

\begin{figure}[tbh]
\includegraphics[scale=0.33]{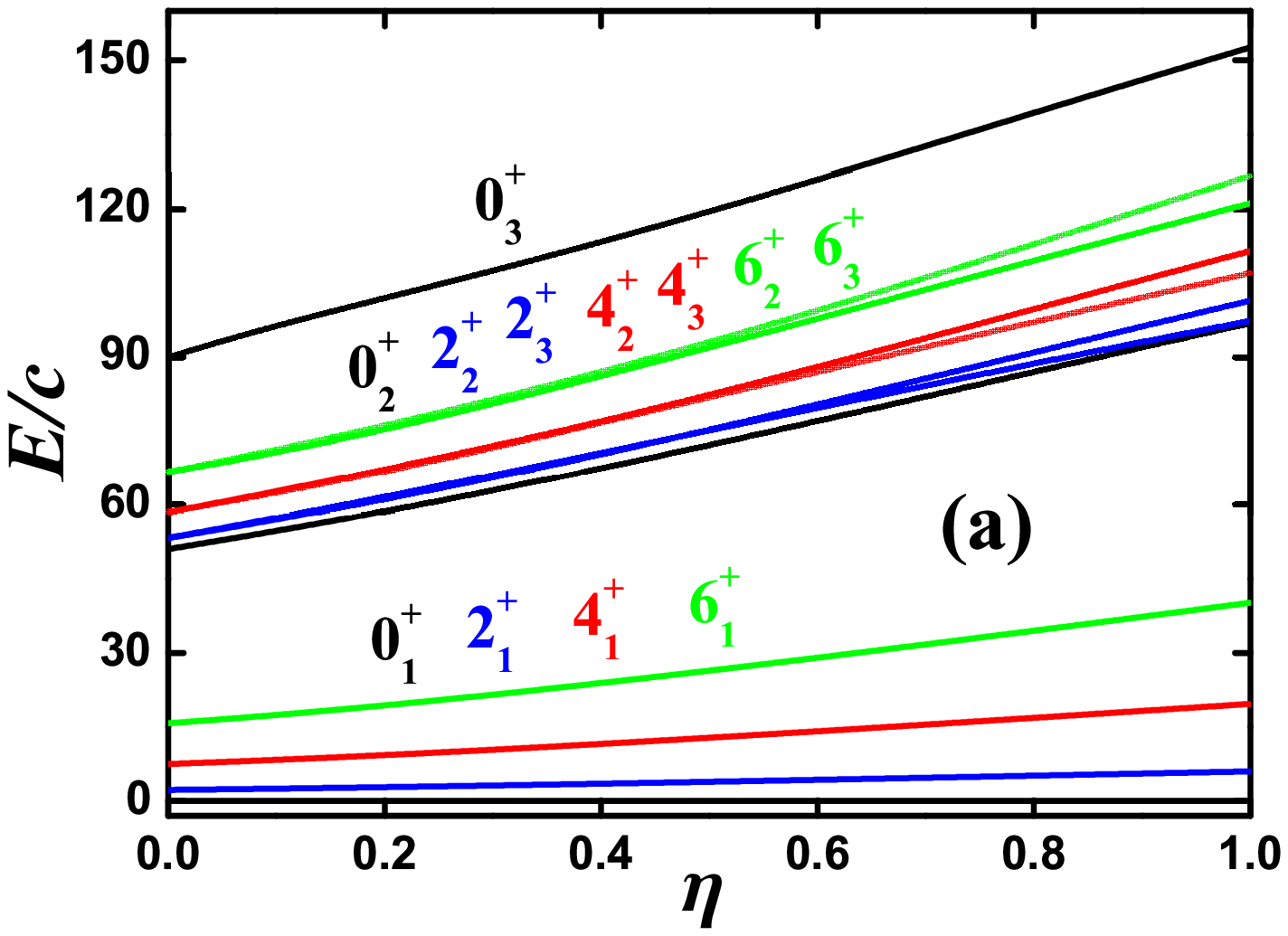}
\includegraphics[scale=0.33]{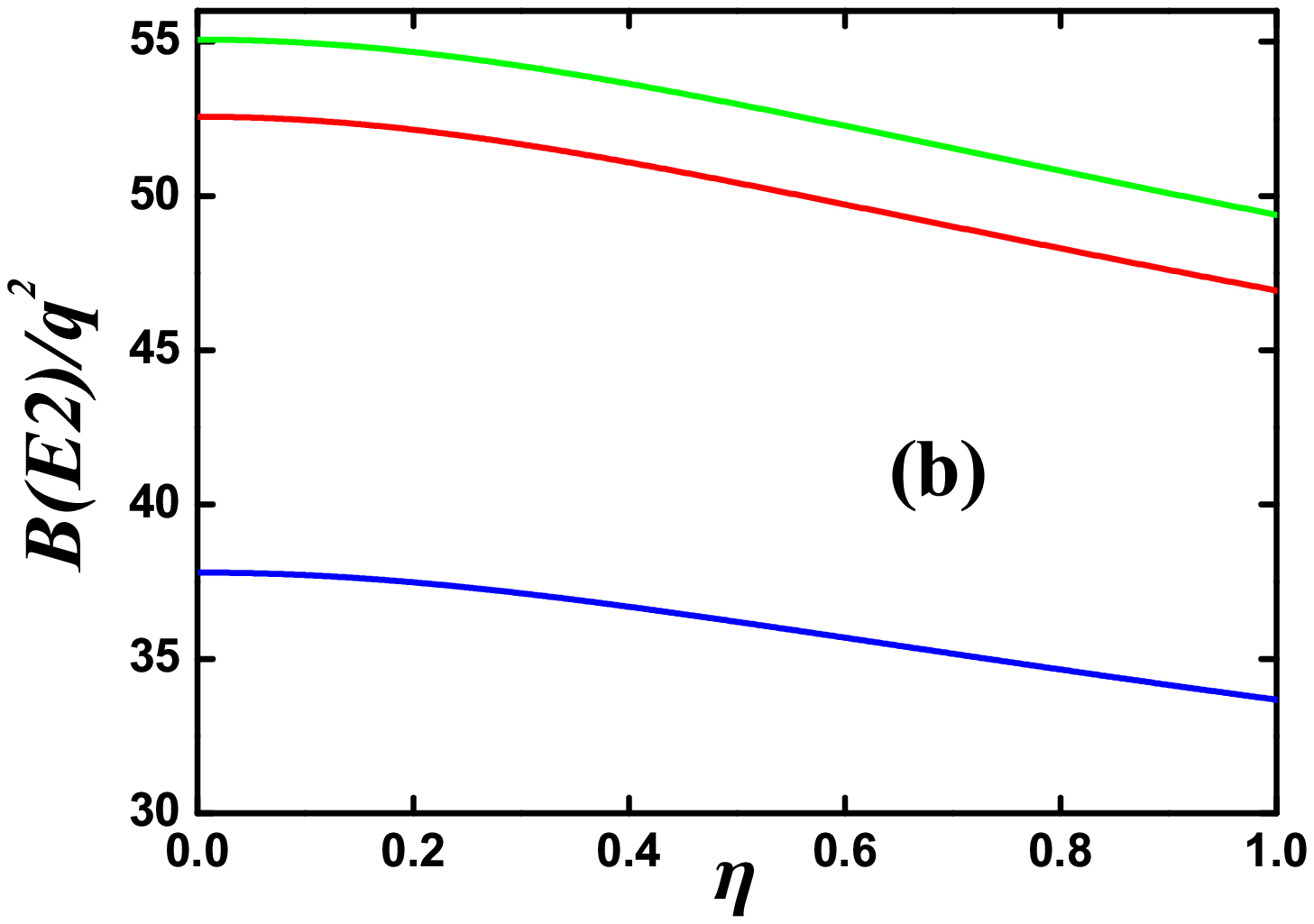}
\caption{(a) The evolutional behaviors of the partial low-lying levels as a function of $\eta$; (b) The evolutional behaviors of the $B(E2; 2_{1}^{+}\rightarrow 0_{1}^{+})$ (blue line) , $B(E2; 4_{1}^{+}\rightarrow 2_{1}^{+})$ (red line), $B(E2; 6_{1}^{+}\rightarrow 4_{1}^{+})$ (green line) as a function of $\eta$.}
\end{figure}

In \cite{Wang20}, due to level-crossing of the $4_{1}^{+}$ and one other $4^{+}$ state in the SU(3) symmetry limit, the $B_{4/2}$ value can be 0. When the $d$ boson number operator $\hat{n}_{d}$ is added, the experimental $B_{4/2}$ value can be obtained. From the analysis of \cite{Pan24}, we know that, if the parameter $\chi$ in $\hat{Q}_{\chi}$ changes from the SU(3) symmetry limit to the O(6) symmetry limit, the anomalous experimental value $B_{4/2}$ may be also obtained. Now we expand the discussions in \cite{Wang20} with the new mechanism.

\begin{figure}[tbh]
\includegraphics[scale=0.33]{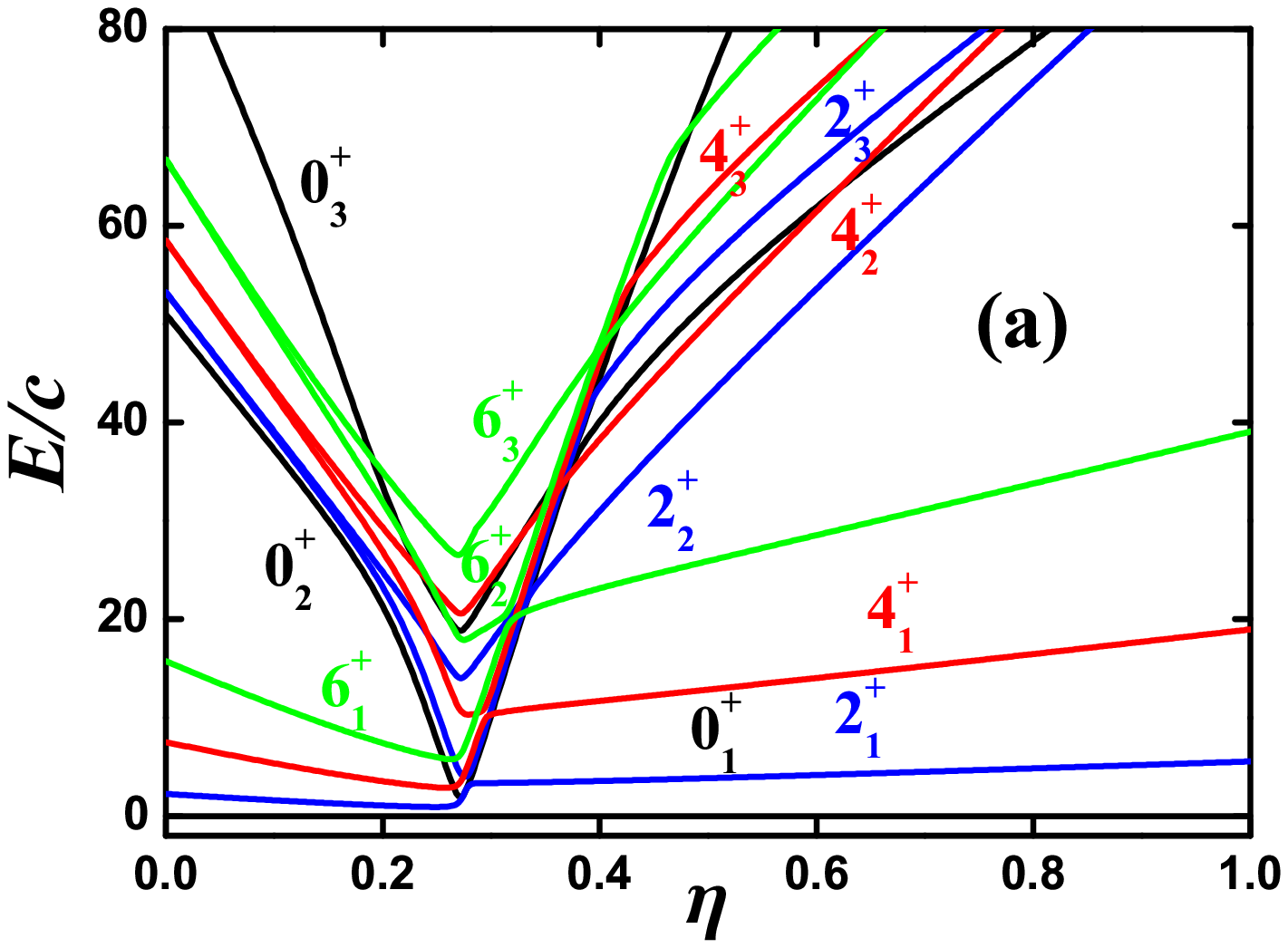}
\includegraphics[scale=0.33]{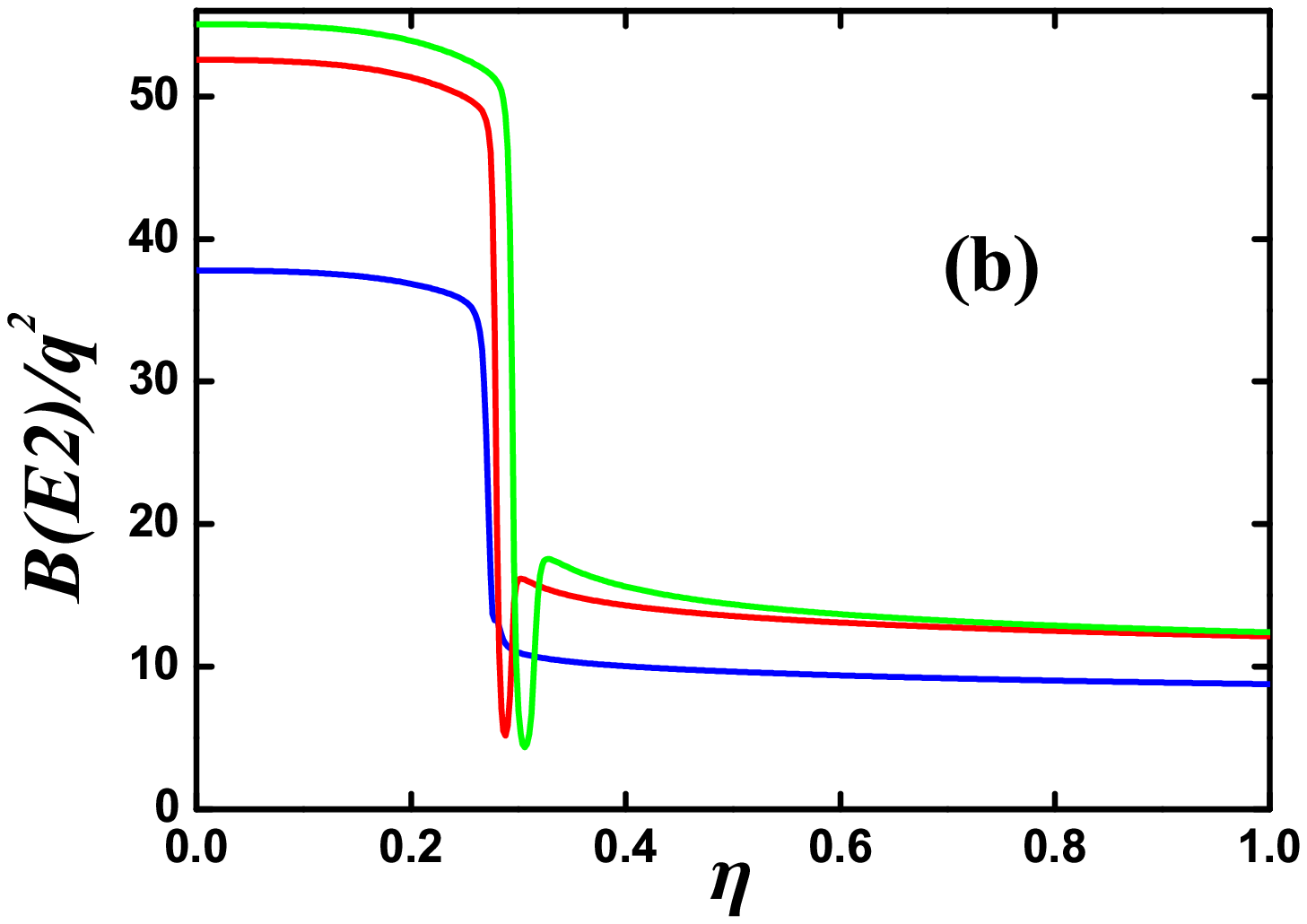}
\caption{(a) The evolutional behaviors of the partial low-lying levels as a function of $\eta$; (b) The evolutional behaviors of the $B(E2; 2_{1}^{+}\rightarrow 0_{1}^{+})$ (blue line) , $B(E2; 4_{1}^{+}\rightarrow 2_{1}^{+})$ (red line), $B(E2; 6_{1}^{+}\rightarrow 4_{1}^{+})$ (green line) as a function of $\eta$.}
\end{figure}

The parameters in \cite{Wang20} are used ($\kappa=60.18$ keV, $\zeta=-9.9653$ keV, $\eta=-17.7193$ keV and $f=-4.515$ keV) and let $\varepsilon_{d}=0$. In the SU(3) symmetry limit, the B(E2) anomaly exists for $B(E2;4_{1}^{+}\rightarrow 2_{1}^{+})=0$. The SU(3) analysis for this case can be found in \cite{Cheng25}. Fig. 7(a) shows the evolutional behaviors of the low-lying levels when the parameter $\chi$ changes from $-\frac{\sqrt{7}}{2}$ to 0. Clearly, shape phase transition arises. At the SU(3) symmetry side, the ground state is really the prolate shape, but its excitations are not similar to the prolate spectra. At the O(6) symmetry side, the sign of the O(6) third-order interactions is negative, so it is an oblate shape \cite{Wangtao}. level-anticrossing of the $2_{1}^{+}$ and $2_{2}^{+}$ states can be easily found. $4_{1}^{+}$, $4_{2}^{+}$ or $6_{1}^{+}$, $6_{2}^{+}$ have no similar phenomena for level-crossing of these levels has occurred in the SU(3) symmetry limit when the parameter of the third-order interaction $[\hat{L}\times \hat{Q} \times \hat{L}]^{(0)}$ decreases \cite{Cheng25}. In Fig. 7(b), B(E2) anomaly exists for $\chi< -1.0$. The left side of the dashed line is for $B_{4/2}\leq0.38$. This is similar to the Fig. 6(b). In section VI, this case will be used to fit the $^{170}$Os. In most part of the region $-\frac{\sqrt{7}}{2}< \chi <-1.0$, the value of $B(E2;6_{1}^{+}\rightarrow 4_{1}^{+})$ is larger than the one of $B(E2;4_{1}^{+}\rightarrow 2_{1}^{+})$.

\begin{figure}[tbh]
\includegraphics[scale=0.33]{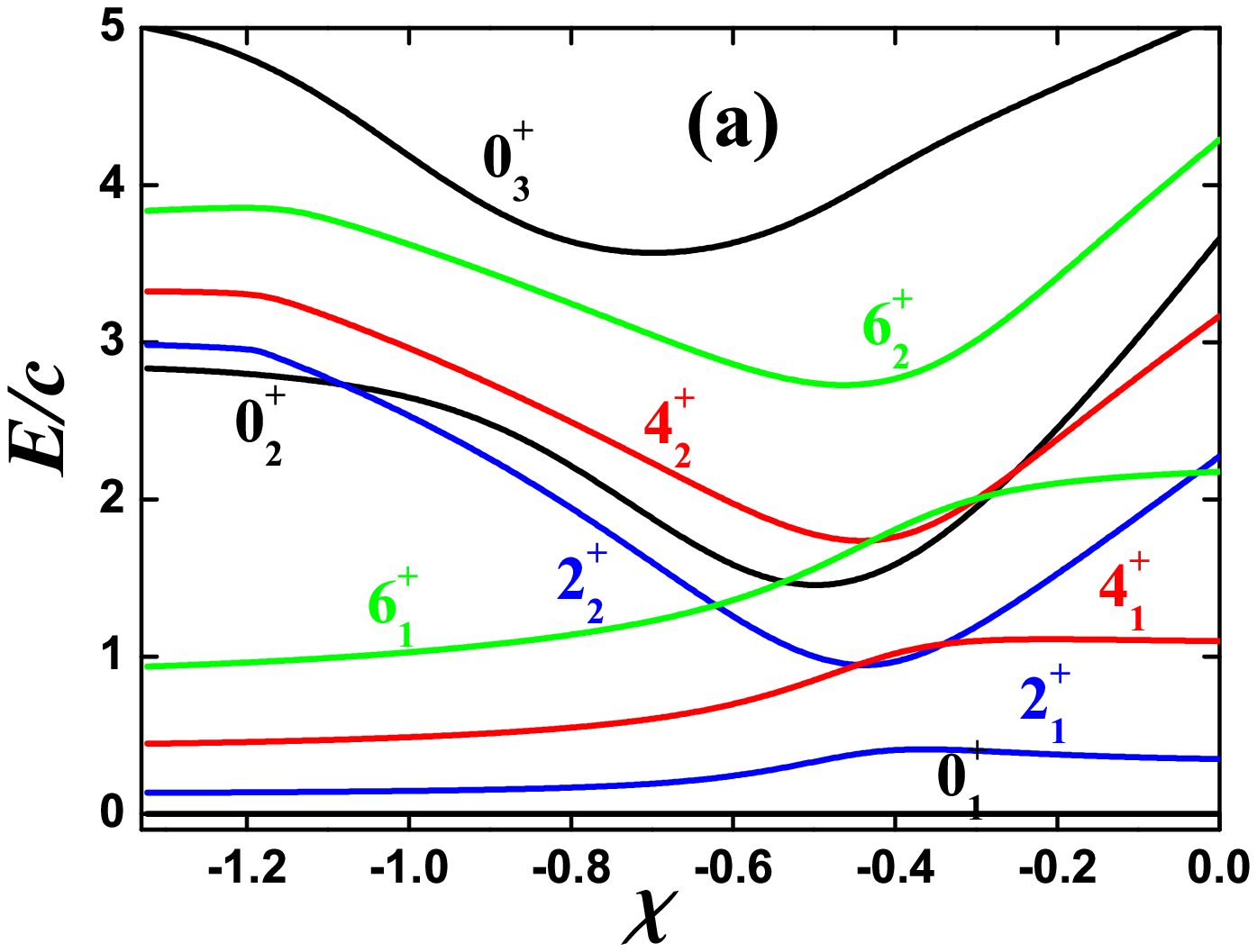}
\includegraphics[scale=0.33]{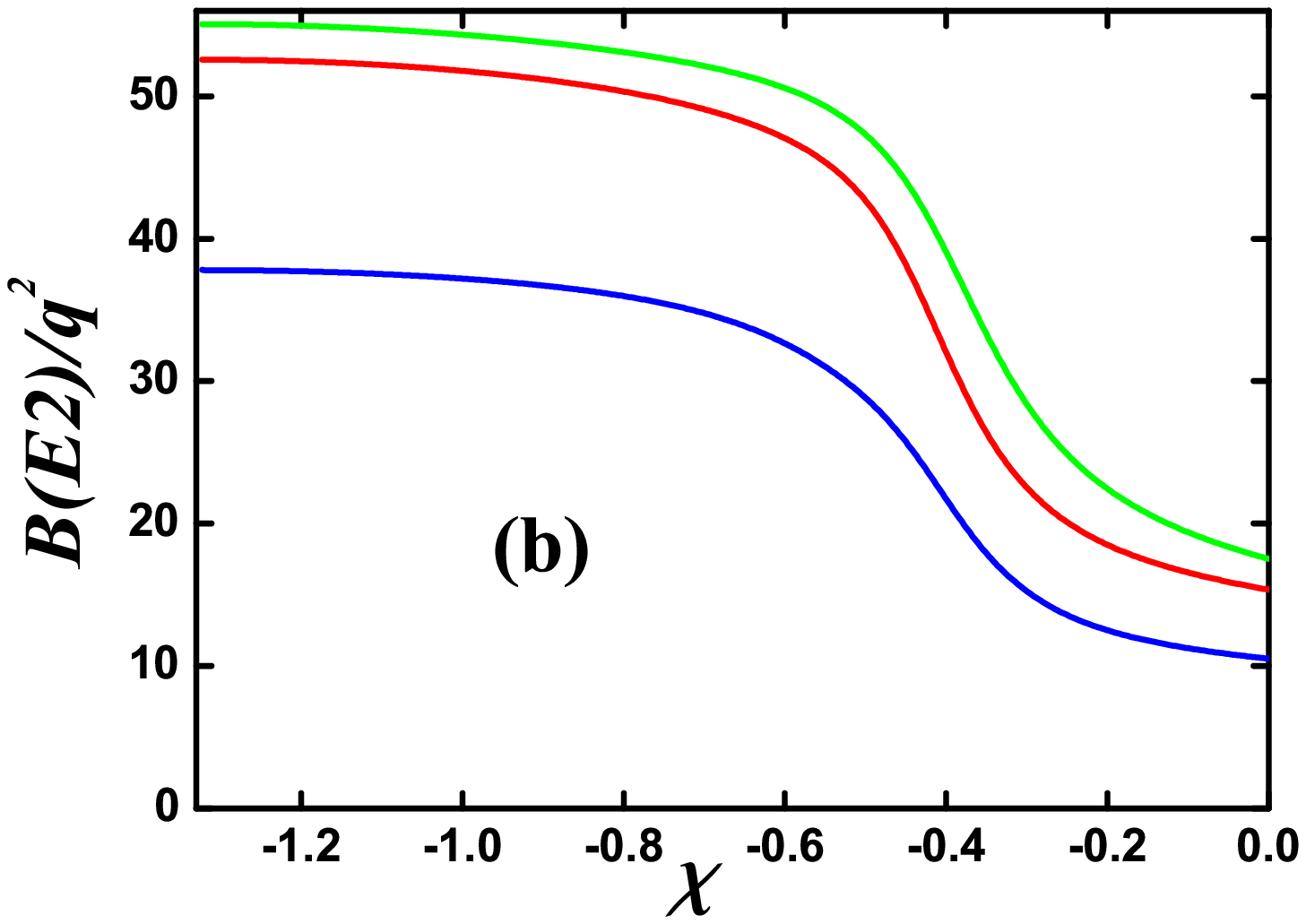}
\caption{(a) The evolutional behaviors of the partial low-lying levels as a function of $\eta$; (b) The evolutional behaviors of the $B(E2; 2_{1}^{+}\rightarrow 0_{1}^{+})$ (blue line) , $B(E2; 4_{1}^{+}\rightarrow 2_{1}^{+})$ (red line), $B(E2; 6_{1}^{+}\rightarrow 4_{1}^{+})$ (green line) as a function of $\eta$.}
\end{figure}

If increasing the parameter of $[\hat{L} \times \hat{Q}_{\chi} \times \hat{L}]^{(0)}$, in the SU(3) analysis, level-crossing of the $4_{1}^{+}$, $4_{2}^{+}$ states will not occur. We choose $\eta=-12.5293$ keV, and other parameters are the same as the ones in Fig. 7(a). In the SU(3) symmetry limit, the B(E2) anomaly $B_{4/2}=0$ does not exist, and the $B_{4/2}$ value is larger than 1.0. Fig. 8(a) shows the evolutional behaviors of the low-lying levels when the parameter $\chi$ changes from $-\frac{\sqrt{7}}{2}$ to 0. Clearly, level-anticrossing of the $4_{1}^{+}$, $4_{2}^{+}$ states can be found. In Fig. 8(b), the B(E2) anomaly can be seen in the parameter region from the SU(3) limit to the O(6) symmetry limit. The parameter region between the two dashed line is for $B_{4/2}\leq0.38$. This case is similar to the Fig. 5(b), and will be also used to discuss the B(E2) anomaly in $^{190}$Os.

If further increasing the parameter of $[\hat{L} \times \hat{Q}_{\chi} \times \hat{L}]^{(0)}$, from the SU(3) symmetry limit and the O(6) symmetry limit, level-anticrossing of the $4_{1}^{+}$, $4_{2}^{+}$ states and the $6_{1}^{+}$, $6_{2}^{+}$ states can be both observed, which is similar to the result in Fig. 4(b) and not shown here.

\begin{figure}[tbh]
\includegraphics[scale=0.33]{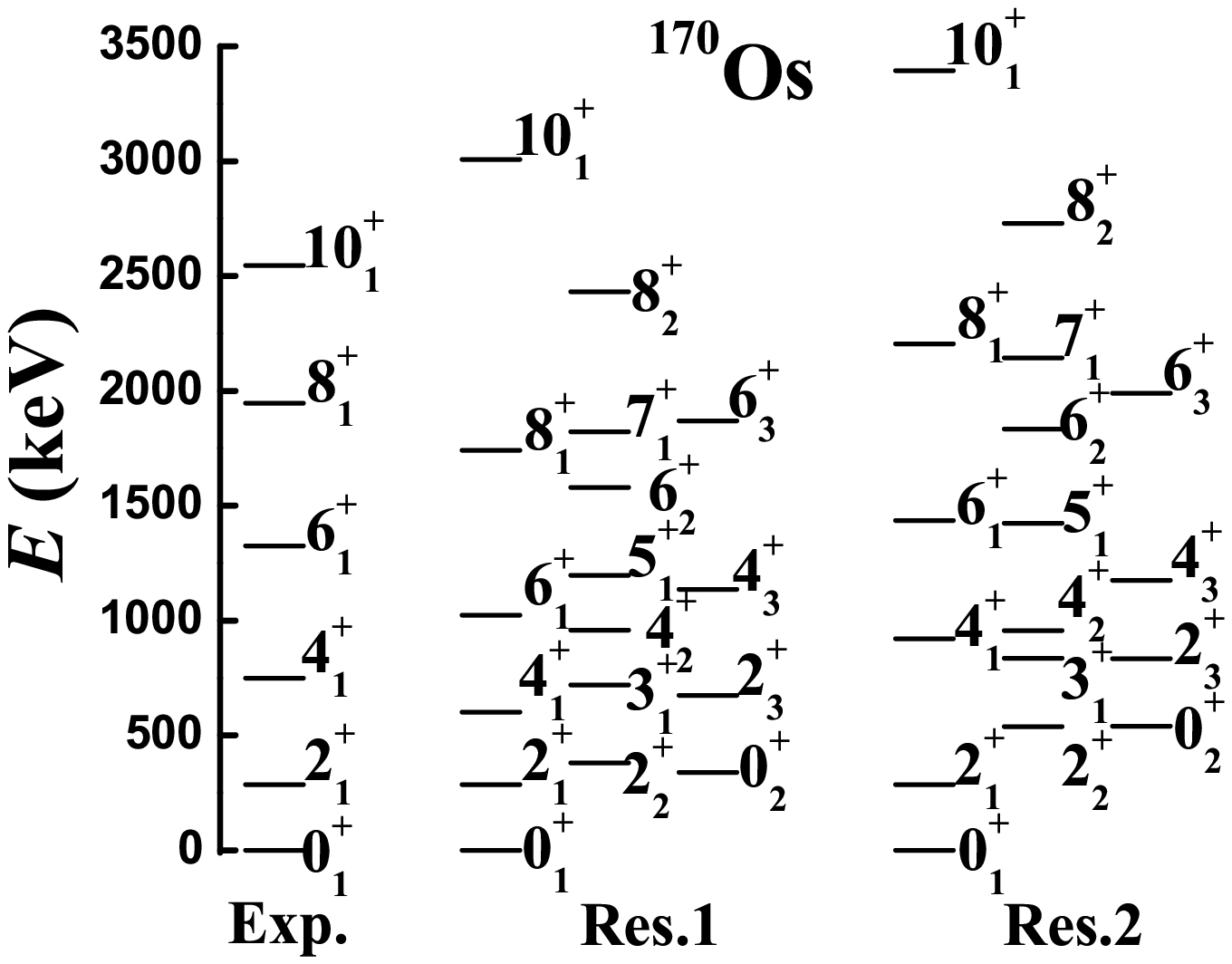}
\caption{The experimental data of the yrast band in $^{170}$Os and the fitting results 1 and 2.}
\end{figure}

Thus the new mechanism in \cite{Pan24} can be combined into \cite{Wang20}. If level-crossing in the SU(3) symmetry limit arises, level-anticrossing from the SU(3) symmetry limit to the O(6) symmetry limit can not occur. On the contrary, it is also true. The newly discovered relationship is universal, and may be a universal mechanism for the emergence of the B(E2) anomaly. The key finding is that, compared to the $\hat{H}_{5}$, the $\hat{H}_{7}$ make the prolate-oblate asymmetric phase transition more prominent.

\section{Some discussions on $[\hat{Q}_{\chi} \times \hat{Q}_{\chi} \times \hat{Q}_{\chi}]^{(0)}$}

In $\hat{H}_{7}$, one may wonder whether this third-order interaction $[\hat{Q}_{\chi} \times \hat{Q}_{\chi} \times \hat{Q}_{\chi}]^{(0)}$ affects the B(E2) anomaly, which needs to be discussed here, and some interesting results are found.

In 1999, Isacker discussed the O(6) third-order interaction $[\hat{Q}_{0} \times \hat{Q}_{0} \times \hat{Q}_{0}]^{(0)}$ \cite{Isacker99}. This work and the discussions on the SU(3) correspondence of the rigid triaxial rotor \cite{Isacker00} inspire the emergence of the SU3-IBM. The O(6) third-order interaction has O(6) symmetry but the O(5) symmetry in the O(6) symmetry reduction chain is broken. Isacker showed that this interaction can describe the prolate shape and its rotation (see Fig. 9), which can replace the SU(3) second-order interaction $-\hat{Q}\cdot \hat{Q}$. Here we further study the relationship between the SU(3) interaction and the O(6) third-order interaction, which can be described by the Hamiltonian
\begin{equation}
\hat{H}_{8}=c[-(1-\eta)\hat{Q}\cdot \hat{Q}+\eta[\hat{Q}_{0} \times \hat{Q}_{0} \times \hat{Q}_{0}]^{(0)}].
\end{equation}
Fig. 10(a) shows the evolutional behaviors of the low-lying levels when the parameter $\eta$ changes from 0 to 1 for $N=9$ (see the green line in Fig. 9). Clearly, the levels evolve in a linear-like way. This result seems very interesting and confusing. At the SU(3) symmetry side, the levels have good quantum number $(\lambda,\mu)$, but at the O(6) symmetry side, only the O(6) quantum number $\sigma$ is used. Fig. 10(b) shows the evolutional behaviors of the E2 transitional rates $B(E2;2_{1}^{+}\rightarrow 0_{1}^{+})$, $B(E2;4_{1}^{+}\rightarrow 2_{1}^{+})$ and $B(E2;6_{1}^{+}\rightarrow 4_{1}^{+})$, the results are similar for any $\eta$. So here, the prolate rotational spectra described by the parameters along the green line in Fig. 9 are very similar, the emergence of the B(E2) anomaly should be found on the upper side of the green line.

In the O(6) symmetry limit, the $[\hat{Q}_{0} \times \hat{Q}_{0} \times \hat{Q}_{0}]^{(0)}$ presents the prolate shape while the $-[\hat{Q}_{0} \times \hat{Q}_{0} \times \hat{Q}_{0}]^{(0)}$ presents the oblate shape \cite{Fortunato11}. In \cite{Wangtao}, it was found that the spectra of the two shapes are also the same. Now we have such a Hamiltonian
\begin{equation}
\hat{H}_{9}=c[-(1-\eta)\hat{Q}\cdot \hat{Q}-\eta[\hat{Q}_{0} \times \hat{Q}_{0} \times \hat{Q}_{0}]^{(0)}],
\end{equation}
which presents the prolate-oblate asymmetric shape phase transition. In Fig. 9, it is shown that, any evolution path through the dashed blue line will generate the prolate-oblate shape phase transition, and $\hat{H}_{7}$ is a typical example. Fig. 11(a) shows the evolutional behaviors of the low-lying levels when the parameter $\eta$ changes from 0 to 1 (see the red line in Fig. 9). Prominent quantum phase transition can be observed, and importantly level-anticrossing between the $2_{1}^{+}$, $2_{2}^{+}$ states, or between the $4_{1}^{+}$, $4_{2}^{+}$ states, or between the $6_{1}^{+}$, $6_{2}^{+}$ states, can be found. In Fig. 11(b), the B(E2) anomaly appears in a narrow parameter region. However this anomaly can not explain the experimental data. When evolving through the dashed blue line, this kind of level-anticrossing can be found at the SU(3) symmetry side, but it occurs only within a narrow parameter region and not useful.

We also consider a prolate-oblate shape phase transition arising at the O(6) symmetry side of the dashed blue line. The evolutional path of the red dashed line in Fig. 9 is studied. Thus we have the Hamiltonian
\begin{equation}
\hat{H}_{9}=c[-\frac{1}{N}\hat{Q}_{\chi}\cdot \hat{Q}_{\chi}-\frac{\kappa}{N^{2}}[\hat{Q}_{\chi} \times \hat{Q}_{\chi} \times \hat{Q}_{\chi}]^{(0)}],
\end{equation}
From the $-\hat{Q}\cdot \hat{Q}$ to the $-[\hat{Q} \times \hat{Q} \times \hat{Q}]^{(0)}$, the prolate-oblate shape phase transition can occur and there is a SU(3) degenerate point at $\kappa_{0}=\frac{\sqrt{35}N}{3(2N+3)}$ \cite{Zhang12,Wang22}. In Fig. 12, $\kappa=\kappa_{0}/2$ is used. It is shown that level-anticrossing and the B(E2) anomaly can not be found.

Thus the most important interaction is $[\hat{L}\times \hat{Q}_{\chi} \times \hat{L}]^{(0)}$ in $\hat{H}_{7}$ for the emergence of the B(E2) anomaly. However, the third-order interaction $[\hat{Q}_{\chi} \times \hat{Q}_{\chi} \times \hat{Q}_{\chi}]^{(0)}$ is still necessary. In other papers related to the oblate shape \cite{Wang23,WangHg}, we can see that this interaction plays an important role, especially in the SU(3) symmetry limit. The $-[\hat{Q}_{\chi} \times \hat{Q}_{\chi} \times \hat{Q}_{\chi}]^{(0)}$ interaction induces the prolate-oblate asymmetric shape phase transition and the the spherical-like spectra \cite{Wang22}. The $[\hat{L}\times \hat{Q}_{\chi} \times \hat{L}]^{(0)}$ interaction mainly induces the B(E2) anomaly. When fourth-order interactions are considered, it will be more complex and discussed in future (rigid triaxial shapes can be found).

\section{B(E2) anomaly in $^{170}$Os}

\begin{figure}[tbh]
\includegraphics[scale=0.33]{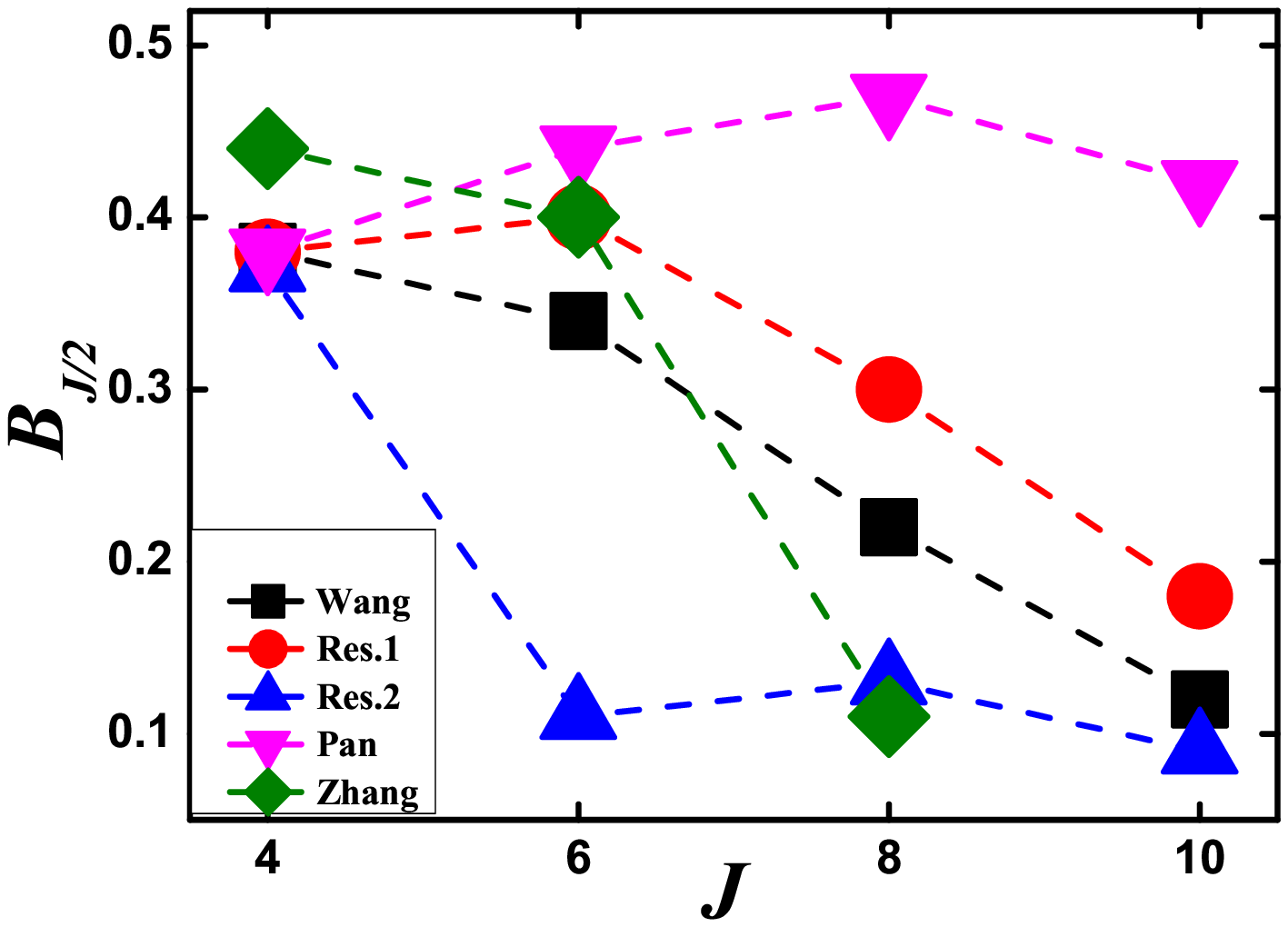}
\caption{$B_{J/2}$ as a function of the parameter $J$. The theories are the fitting results 1 and 2 in this paper, and \cite{Wang20}(Wang), \cite{Pan24}(Pan) and \cite{Zhang25}(Zhang).}
\end{figure}

Now we fit the $^{170}$Os with the Hamiltonian $\hat{H}_{7}$. The fitting parameters are deduced from section IV for the energy of the $2_{1}^{+}$ state is the same as the experimental value. The parameters of the result 1 are $\chi=-1.2$, $\varepsilon_{d}=138$ keV, $\kappa=67.24$ keV, $\zeta=-11.13$ keV, $\eta=-19.8$ keV and $f=-5.04$ keV (deduced from Fig. 7) , and the ones of the result 2 are $\chi=-1.17207$, $\varepsilon_{d}=0$, $\kappa=93.56$ keV, $\zeta=-15.49$ keV, $\eta=-19.47$ keV and $f=-7.02$ keV (deduced from Fig. 8).  Fig. 13 presents the fitting results. These results are similar to the ones in \cite{Wang20,Pan24}.

Defining $B_{J/2}=B(E2;J_{1}^{+}\rightarrow (J-2)_{1}^{+})/B(E2;2_{1}^{+}\rightarrow 0_{1}^{+})$. Fig. 14 presents the $B_{J/2}$ evolution with the parameter $J$. The experiment result of $B_{4/2}$ in $^{170}$Os is 0.38(11), so these theoretical results fit well for $B_{4/2}$. The key difference is $B_{6/2}$. In \cite{Wang20,Zhang25}, the $B_{6/2}$ is smaller than the $B_{4/2}$. In result 2, the $B_{6/2}$ is much smaller than the $B_{4/2}$, which can be also found in \cite{Pan24} but not mentioned. In Fig. 5, the $B_{6/2}$ can reduce to a very small value 0.015 when $B_{4/2}$ is 0.38. Result 1 and \cite{Pan24} show an opposite trend. This needs further experimental confirmation. In \cite{Saygi17}, the $B_{6/2}$ value in $^{166}$W is 0.12(4). In \cite{Zhang22}, the two SU(3) fourth-order interactions are also considered, and the $B_{6/2}$ can reduces to 0.12. In result 2, this small value can be also obtained. In future, it is necessary to discuss the B(E2) anomaly in $^{166}$W, and its relationship with the fourth-order interactions.

\section{Conclusion}

Using the SU(3) analysis technique proposed in the previous paper, we observed that, the new mechanism found in \cite{Pan24} is also related to the SU(3) limit. In this paper, we illustrate the reason for this. We find the third-order interactions $[\hat{L}\times \hat{Q}_{\chi} \times \hat{L}]^{(0)}$ is vital for the emergence of the B(E2) anomaly. A specific relationship is found. If level-crossing occurs in the SU(3) symmetry limit when the parameter of $[\hat{L}\times \hat{Q} \times \hat{L}]^{(0)}$ decreases, level-anticrossing from the SU(3) symmetry limit to the O(6) symmetry limit cannot appear. The opposite result is also true. The two ideas in \cite{Wang20} and \cite{Pan24} can merge into a general explanatory framework, and the B(E2) anomaly in $^{170}$Os is discussed. In future works, the two SU(3) fourth-order interactions are also considered to relate with the level-anticrossing phenomena.

In the analysis of this paper, the SU(3) symmetry limit presents the prolate shape. In \cite{Zhang25}, Y. Zhang \emph{et al.} also proposed a new mechanism in the SU3-IBM, which is related to level-anticrossing and the oblate shape \cite{Cheng25}, which will be discussed in next paper (II) \cite{Tie25}.

\end{document}